\documentclass{acm_proc_article-sp}

\newcommand{\albert}[1]{\textsf{\textcolor{albert_color}{[{\bf Albert says}: #1]}}}

\usepackage{hyperref}
\usepackage{amsmath}
\usepackage{amssymb}
\usepackage{multirow}
\usepackage{rotating}
\usepackage{caption}
\usepackage{color}
\usepackage{enumitem}
\setlistdepth{9}

\definecolor{albert_color}{rgb}{0,0.5,1}
\definecolor{srm_color}{rgb}{0.2,.64,0}

\DeclareGraphicsExtensions{.jpg,.pdf,.png}

\begin{document}

\title{Procedural wood textures} 

\numberofauthors{4}

\author{
\alignauthor
Albert J. Liu \\
\affaddr{Cornell University}
\email{ajul@cs.cornell.edu}
\alignauthor
Stephen R. Marschner \\
\affaddr{Cornell University}
\email{srm@cs.cornell.edu}
\alignauthor
Victoria E. Dye \\
\affaddr{Cornell University}
\email{ved8@cornell.edu}
}

\maketitle

\begin{abstract} 
Existing bidirectional reflectance distribution function (BRDF) models are capable of capturing the distinctive highlights produced by the fibrous nature of wood. However, capturing parameter textures for even a single specimen remains a laborious process requiring specialized equipment. In this paper we take a procedural approach to generating parameters for the wood BSDF. We characterize the elements of trees that are important for the appearance of wood, discuss techniques appropriate for representing those features, and present a complete procedural wood shader capable of reproducing the growth patterns responsible for the distinctive appearance of highly prized ``figured'' wood. Our procedural wood shader is random-access, 3D, modular, and is fast enough to generate a preview for design.
\end{abstract}




\section{Introduction}

Wood is a commonly appearing material in many kinds of graphics scenes, particularly indoor scenes full of wooden floors, cabinets, furniture, etc.  The standard way to model wood is to use high-resolution photographic texture maps to control the diffuse color, a microfacet model for the surface, and a bump map derived from the color maps to introduce surface details.  This approach has several problems, though. It requires a lot of high-resolution textures, because the grain patterns of wood have large-scale structures that defeat tiling and window-based texture synthesis and make any repetition easy to spot.  Coming up with textures for curved surfaces that are carved or turned from solid wood is difficult, since the patterns on the surface are inherently connected to the shape.  But also, representing wood with a diffuse color ignores the distinctive shifts in color and texture on a wood surface depending on view and illumination---an essential part of the beauty of this prized traditional material.

The appearance of a wood surface depends on the internal structure of the board, which in turn reflects the anatomy of the tree it came from, with an intricate arrangement of long, thin cells that reflect light very anisotropically.  Previous work \cite{Marschner:2005:FinishedWood} has shown that the surface reflectance of finished wood can be represented well by a model that includes a diffuse component and a separately colored fiber-reflection component that is controlled by a direction texture giving the 3D direction of fibers at every point on the surface.  These color and direction textures can be acquired for a particular board by a process using many images under controlled illumination, but using this approach to capture enough texture to cover large areas without repeating is daunting.  Also, data captured for flat surfaces will not work for curved surfaces carved from solid wood.

An alternative approach is to use a solid texture \cite{Peachey:1985:SolidTexturing}, which makes carving a surface out of a volume easy.  Volume texturing has the advantage of not requiring a surface parameterization, and it automatically allows the wood to be cut along different planes to create different grain patterns. Since there is no simple way to measure a solid texture for a block of wood, solid textures are generally defined procedurally, and indeed, wood grain is a classic application for procedural 3D textures. However, previous work on procedural wood textures has only modulated diffuse color, precluding the simulation of directional illumination effects.

The aims of this paper are as follows:
\begin{itemize}
	\item Characterize the specific elements of wood anatomy that are important for the visual appearance of wood.
	\item Apply noise and distortion to represent such elements of wood, including the fiber directions necessary for replicating the patterns of anomalous growth that lead to highly prized ``figured'' lumber.
	\item Show how these methods can be combined into a complete procedural shader that can simulate a wide variety of real woods.
\end{itemize}

We have chosen to restrict ourselves to a \textit{random-access} model as defined in e.g. \cite{Lefebvre:2007:RuntimeTextureSynthesis}: the model may be evaluated at any location in constant time regardless of what points have been previously evaluated. In other words, our model is based on closed-form methods rather than taking a simulational or progressive patch approach. The random-access approach allows a nearly arbitrary amount of texture to be generated using a relatively small set of global parameters. If greater performance is desired, conventional precomputed textures can easily be generated.

In the following sections we discuss related work on textures and surface reflection and the relevant facts about the anatomy and growth patterns of trees, then discuss the building blocks of our model---impulses, noise, and warps---and how those blocks are assembled into a model that has the flexibility to convey the appearance of a wide range of different wood species. Our approach is based on random collections of impulses in 3D space, which are used both to instantiate discrete anatomical features and to define noise functions that are used to modulate color and to generate 3D warps.

\section{Related work}

Our model builds directly on previous work in wood appearance and procedural noise.

\subsection{Measuring and Modeling the Appearance of Finished Wood} \label{sec:fiber_specular}
The BSDF of our model is that of \cite{Marschner:2005:FinishedWood}. This BSDF is the sum of a Lambertian diffuse component and a specular component. For the specular component the BSDF models the reflection from subsurface fibers in much the same way reflection is modeled for hair and fur. The effect is that incoming light is scattered into a cone with some spread. Specifically, the reflectance behavior post-refraction is given by

\begin{align}
	f_f \left( \mathbf{u}, \mathbf{v}_i, \mathbf{v}_r \right) 
	= k_f \frac{g \left( \beta, \psi_h \right)}{\left( \cos \left( \psi_d / 2 \right) \right)^2}
\end{align}

where $\mathbf{u}$ is the fiber direction, $k_f$ is the fiber reflection color, $g \left(\sigma, x\right)$ is a normalized Gaussian, $\mathbf{v}_i, \mathbf{v}_r$ are the incident and reflected directions, $\beta$ is the highlight width (typically around 10 to 15 degrees), and

\begin{align}
	\psi_i &= \arcsin \left( \mathbf{v}_i \cdot \mathbf{u} \right)
	& \psi_d &= \psi_r - \psi_i \\
	\psi_r &= \arcsin \left( \mathbf{v}_r \cdot \mathbf{u} \right)
	& \psi_h &= \psi_r + \psi_i
\end{align}

The effect of the surface interface's change in index of refraction is applied on top of this by adjusting the incoming and outgoing angles and scaling by the appropriate Fresnel transmission factors.

This BSDF is able to capture effects seen in real wood that cannot be reproduced by surface-only anisotropy such as the Ward model \cite{Ward:1992:Anisotropic}. However, acquiring parameter data has long been a practical problem in using this wood BSDF. The original paper uses a specialized fitting process which requires specific equipment and a large number of photographs per sample. The primary purpose of our model is to address these difficulties in using the wood BSDF.

\subsection{Exemplar-based methods}

\begin{figure}
\centering
\includegraphics[width=0.9\linewidth]{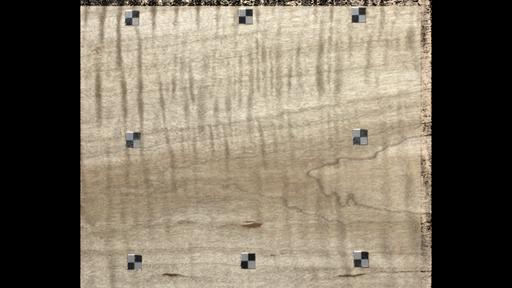} \\
\includegraphics[width=0.45\linewidth]{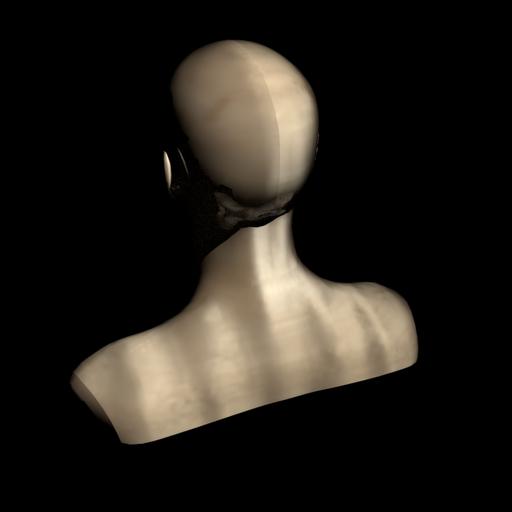}
\includegraphics[width=0.45\linewidth]{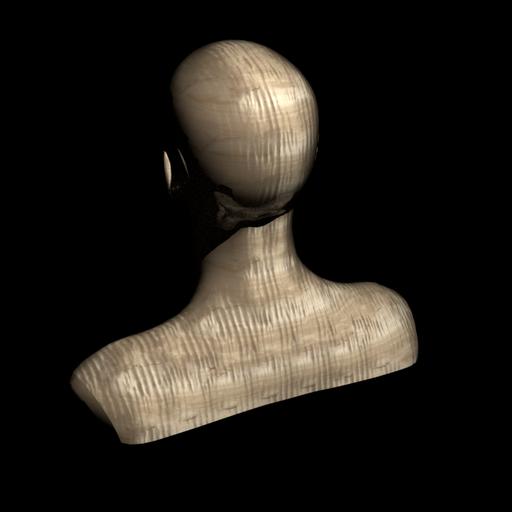}
\caption{\textbf{Top:} Rendering produced using the 650x515 original texture set from \cite{Marschner:2005:FinishedWood}. \textbf{Left:} Non-tiled rendering using the 128x128x128 solid texture set from \cite{Kopf:2007:SolidFrom2D}. \textbf{Right:} Tiling the texture 10 times per side using the same solid texture set results in a sharper appearance at the cost of repetition artifacts. Furthermore, the result lacks the types of global correlations found in real wood such as (approximately) cylindrical growth ring surfaces. Model by Lee Perry-Smith, Morgan McGuire, and Guedis Cardenas; downloaded from Morgan McGuire's Computer Graphics Archive \url{http://graphics.cs.williams.edu/data}.}
\label{fig:kopf}
\end{figure}

Exemplar-based methods (as reviewed in e.g. \cite{Wei:2009:STARExampleBased}) have been successful in generating a larger amount of texture from a single examplar and are applicable to a wide varity of exemplars. Works such as \cite{Kopf:2007:SolidFrom2D} are even capable of generating a 3D texture from 2D examplar(s). However, they have traditionally lacked the advantages of compact representation, random access, easy editability, and infinite resolution of procedural methods. While exemplar-based methods have advanced (again as reviewed in \cite{Wei:2009:STARExampleBased}) in each of these areas, procedural methods inherently offer all of these advantages in concert. Furthermore, exemplar-based methods typically rely on limited windows, which limits their ability to reproduce long-range correlations such as the distinctive changes in the shape and width of growth rings as the slicing angle changes from tangential to radial. Figure \ref{fig:kopf} shows an example generated using \cite{Kopf:2007:SolidFrom2D}.


\subsection{Noise functions}
A key component of most procedural textures is noise functions. A survey is given in \cite{Lagae:2010:SurveyProceduralNoise}. Excessive regularity in a texture looks unnatural and jarring, so  introducing (pseudo-) randomness into a procedural texture model can therefore produce pleasing variation in the resulting texture. Perhaps the most famous example of noise is Perlin noise \cite{Perlin:1985:PerlinNoise,Perlin:2002:ImprovingNoise}. Other types of noise include cellular noise \cite{Worley:1996:CellNoise} and wavelet noise \cite{Cook:2005:WaveletNoise}. There is also the family of sparse convolution noises, which convolve a set of impulses with a kernel, first introduced by (\cite{Lewis:1984:TextureSynthesis,Lewis:1989:AlgorithmsForSolidNoise}) and further developed by \cite{vanWijk:1991:SpotNoise}. More recently, the Gabor kernel has emerged as an attractive choice \cite{Lagae:2009:SparseGabor,Lagae:2011:ImprovingGabor,Galerne:2012:GaborExample} due to its excellent spatial and spectral properties. We have chosen sparse convolution noise here. For simplicity, we use a white noise (uniform) distribution of impulses, though blue noise distributions such as \cite{Lagae:2005:ObjectDistribution,Lagae:2008:TileMethods} are an option to produce less ``clumping''.

\section{The elements of wood}

\begin{figure}
\centering
\includegraphics[width=0.75\linewidth]{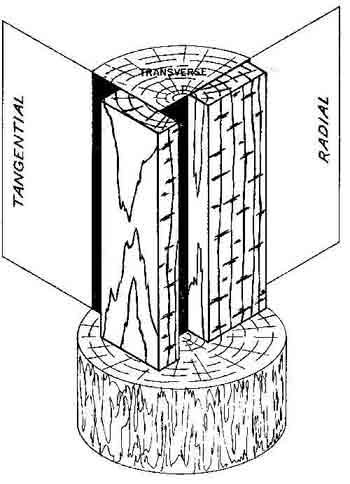}
\caption{The principal cutting planes of wood. Diagram courtesy of Alabama Agricultural Experiment Station \cite{Beals:1977:FigureInWood}.}
\label{fig:planes}
\end{figure}

Wood comes from the cambium of trees, including the hardwoods and softwoods.  Our model is organized around the anatomy of wood, so we begin with some discussion of the relevant features (See Figure~\ref{fig:planes} for an illustration of planes used to describe these features).  We will focus on hardwoods (trees belonging to the angiosperms---not always harder wood) because they contain a greater variety of anatomical structures than softwoods and also contain nearly all woods that are prized for their appearance.  Softwoods generally contain a subset of the same features and can also be handled by our model, but other woody plants used for lumber (such as bamboo) are entirely different in structure and are not within the intended scope. 
\cite{Hoadley:1980:UnderstadingWood,Panshin:1970:TextbookOfWood}

\begin{figure}
\centering
\includegraphics[width=0.9\linewidth]{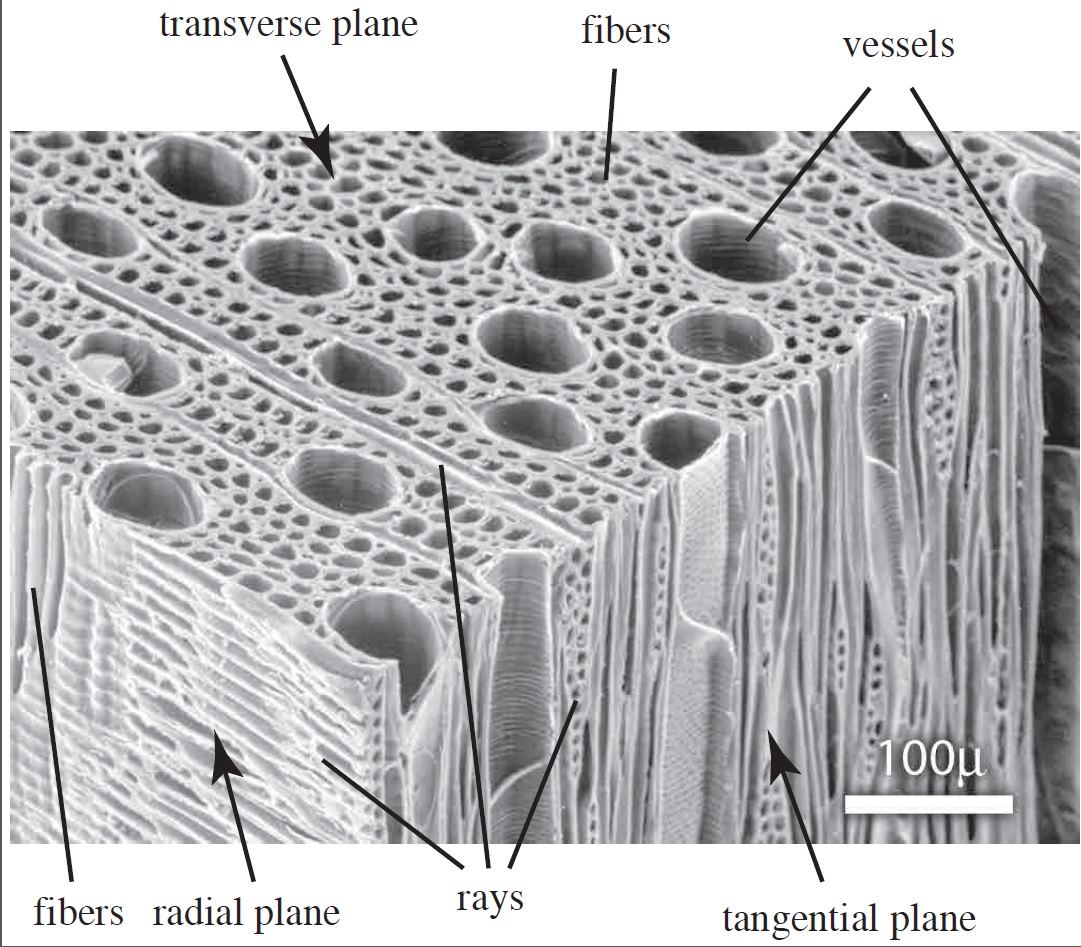}
\caption{Scanning electron micrograph of red maple, showing vessels, rays, and fibers, as well as the radial, tangential, and transverse planes. [NC Brown Center for Ultrastructure Studies, SUNY College of Environmental Science and Forestry, Syracuse, NY]}
\label{fig:labeled-sem}
\end{figure}

\subsection{Seasonal growth}
Perhaps the most obvious feature in the appearance of most woods are growth rings. Growth rings result from the contrast between earlywood, which is produced during the spring, grows quickly, tends to be lighter in color, has larger and thinner-walled cells, and sometimes has larger pores (see below), versus latewood, which is produced during the summer and is the opposite. There are some species with indistinct growth rings, such as certain tropical hardwoods that grow in climates without strong seasonal variations.

\subsection{Longitudinal and ray fibers}
The majority of cells (about 90\% by volume depending on species) in wood run in the longitudinal direction. However, there are also ribbon-like cells or clusters of cells which grow outward in the radial direction, called rays. Rays are typically very narrow in the circumferential direction and wider in the longitudinal direction. Depending on species, they may be too small to be seen easily by the naked eye or up to as large about a millimeter thick and a few centimeters tall \cite{Hoadley:1980:UnderstadingWood}.

Visually, rays can produce a striking visual effect since their fibers run in a direction perpendicular to the main mass of fibers, thus produces a different reflective effect on the surface. On the tangential plane rays generally appear dark, whereas on transverse and radial planes they can be very bright for certain illumination configurations.
\subsection{Pores/vessels}

\begin{figure}
\centering
\includegraphics[width=0.3\linewidth]{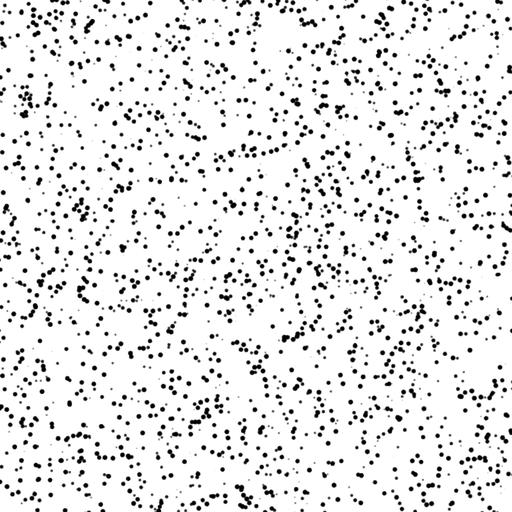} 
\includegraphics[width=0.3\linewidth]{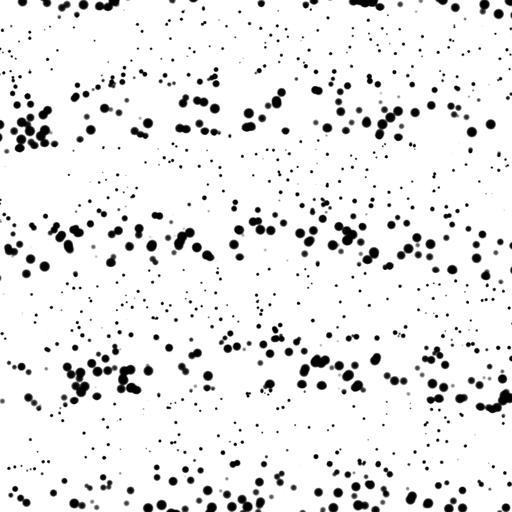} 
\includegraphics[width=0.3\linewidth]{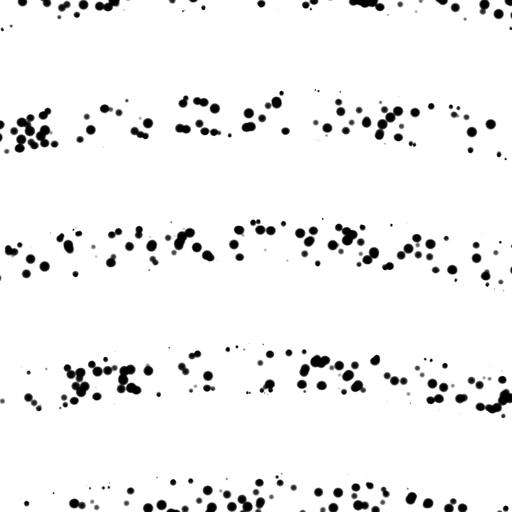} 
\caption{Vessel density may be uniform in diffuse-porous woods (left), only visible in the earlywood in ring-porous woods (right), or somewhere in-between in semi-ring-porous woods (center).}
\label{fig:porous}
\end{figure}

Vessels are hollow longitudinal tubes only found in the so-called hardwoods. When cut, they produce openings or indentations on the surface of the wood referred to as pores. Since wood is usually cut so that the longitudinal direction is in the plane of the surface, pores most often show up as long grooves on the surface of the wood. Unless the pores are filled, such indentations will remain even after the wood is coated. If a staining finish is applied to the wood, the pores will tend to absorb more of the finish, giving them a darker color than the rest of the wood.

The seasonal size and distribution of vessels depends on the species. In species termed diffuse-porous, the size of vessels is independent of the season. At the opposite extreme are ring-porous species, where large pores occupy most of the volume of the earlywood, but the latewood pores are negligibly tiny. In these species the pores rather than the inherent color of the wood may be the primary feature of growth rings. Still other species lie between these two extremes, with larger pores in the earlywood and smaller but still significant pores in the latewood. See Figure \ref{fig:porous} for an illustration.

\subsection{Figure}
Even the main fibers of wood do not grow perfectly longitudinally. Variations in the fiber direction cause distinctive visual effects known as figure.

\begin{figure}
\centering
\includegraphics[height=2in]{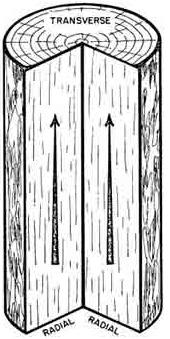}
\includegraphics[height=2in]{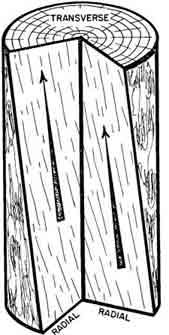}
\includegraphics[height=2in]{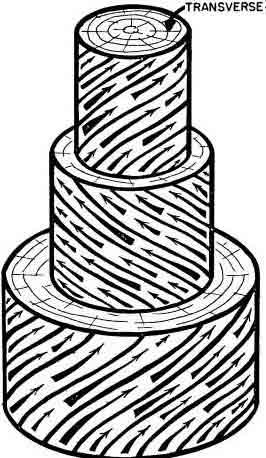}
\caption{\textbf{Left:} In straight-grained woods the longitudinal fibers grow parallel to the trunk axis. \textbf{Center:} In spiral-grained woods the fibers follow a helical path around the axis. \textbf{Right:} The spiral direction may reverse over time resulting in interlocked grain. Diagrams courtesy of Alabama Agricultural Experiment Station \cite{Beals:1977:FigureInWood}.}
\label{fig:grain}
\end{figure}

The overall direction of the fibers in a tree may be parallel to the axis in what is termed straight grain. However, in some woods the fibers may instead follow a helical path around the axis rather than growing parallel to it. This is called spiral grain. In some specimens the direction of the spiral may even reverse itself over time, resulting in interlocked grain. This shows up as alternating light and dark bands on the radial surface known as stripe figure. These types of grain are illustrated in Figure \ref{fig:grain}.

More localized deviations from true longitudinal can be found as well: waves in the fiber can cause deviations in the radial and/or tangential directions (see Figures \ref{fig:distortion_radial} and \ref{fig:distortion_tangential}), which is termed curly or fiddleback figure. Bumps in the radial direction may result in blister, quilted, or birds-eye figure depending on their shape \cite{Beals:1977:FigureInWood}.


\section{Sparse convolution noise}

We now move on to discuss the core techniques we use to represent these anatomical features of wood.

Regularities in a texture can give it a too-perfect, unnatural appearance, and indeed real wood exhibits a wide degree of variation in all aspects even within the same species or tree, whether due to differences in genetics, climate, sunlight, damage, or other causes. In order to introduce some variation to our textures, we use the usual technique of basing parts of our texture on pseudorandom noise rather than only ``purely'' deterministic structure.

We have selected sparse convolution noise as our family of noise functions. Sparse convolution noise consists of a set of impulses (hence ``sparse'') convolved with a finite-support kernel. It has the advantages of versatility via choice of kernel and  appropriateness for both distributing discrete instances of objects (such as rays and vessels) and less structured phenomena (such as variations in fiber direction). A description of sparse convolution noise and our particular adaptation of it to wood textures follows.

\subsection{Impulse search} To determine the value of the noise at a particular query point, we need to determine the locations of all impulses that may be close enough to the query point to affect the value there. These locations need to be consistent from query to query.

\begin{figure}
\begin{tabular}{cc}
(a) \includegraphics[width=0.4\linewidth]{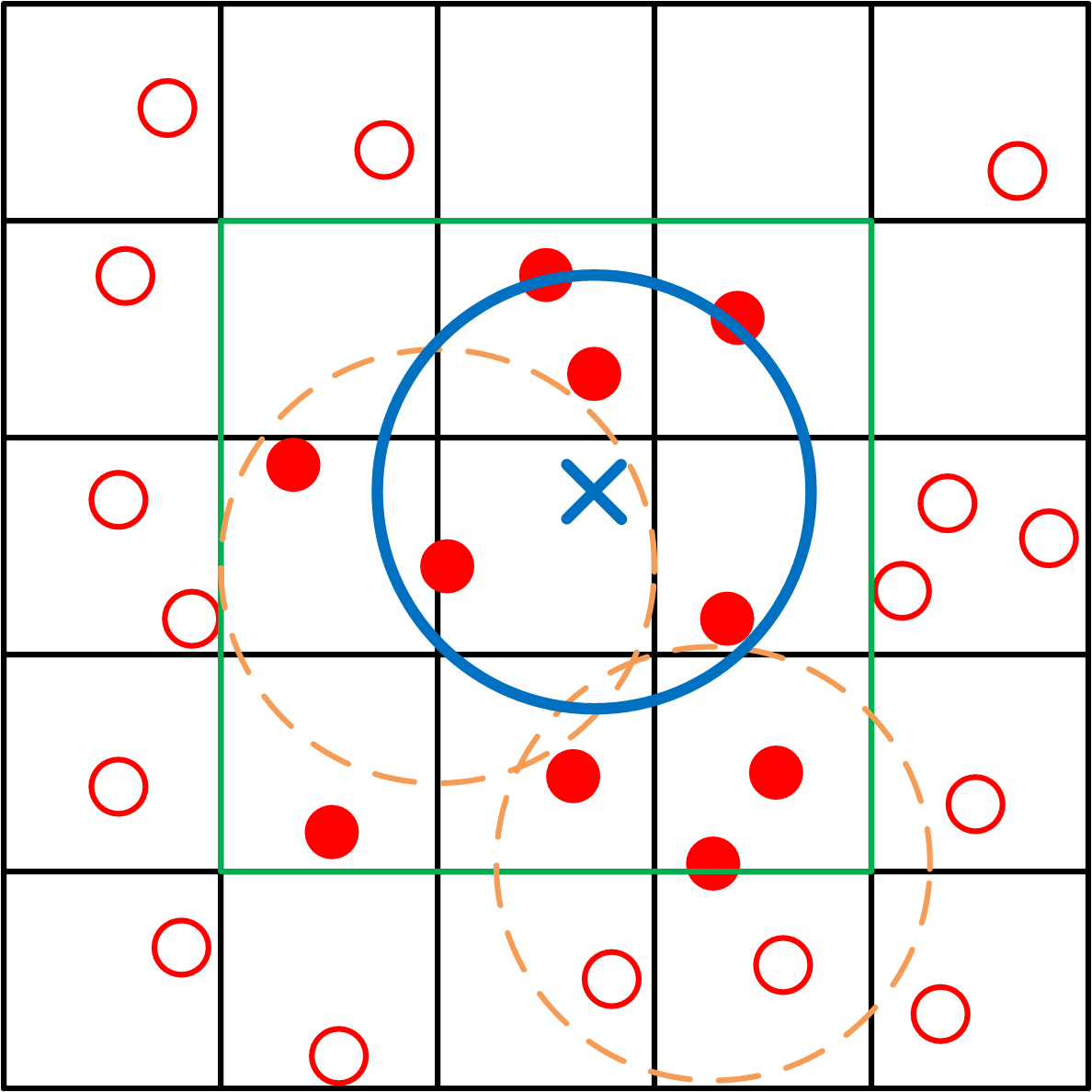} &
(b) \includegraphics[width=0.4\linewidth]{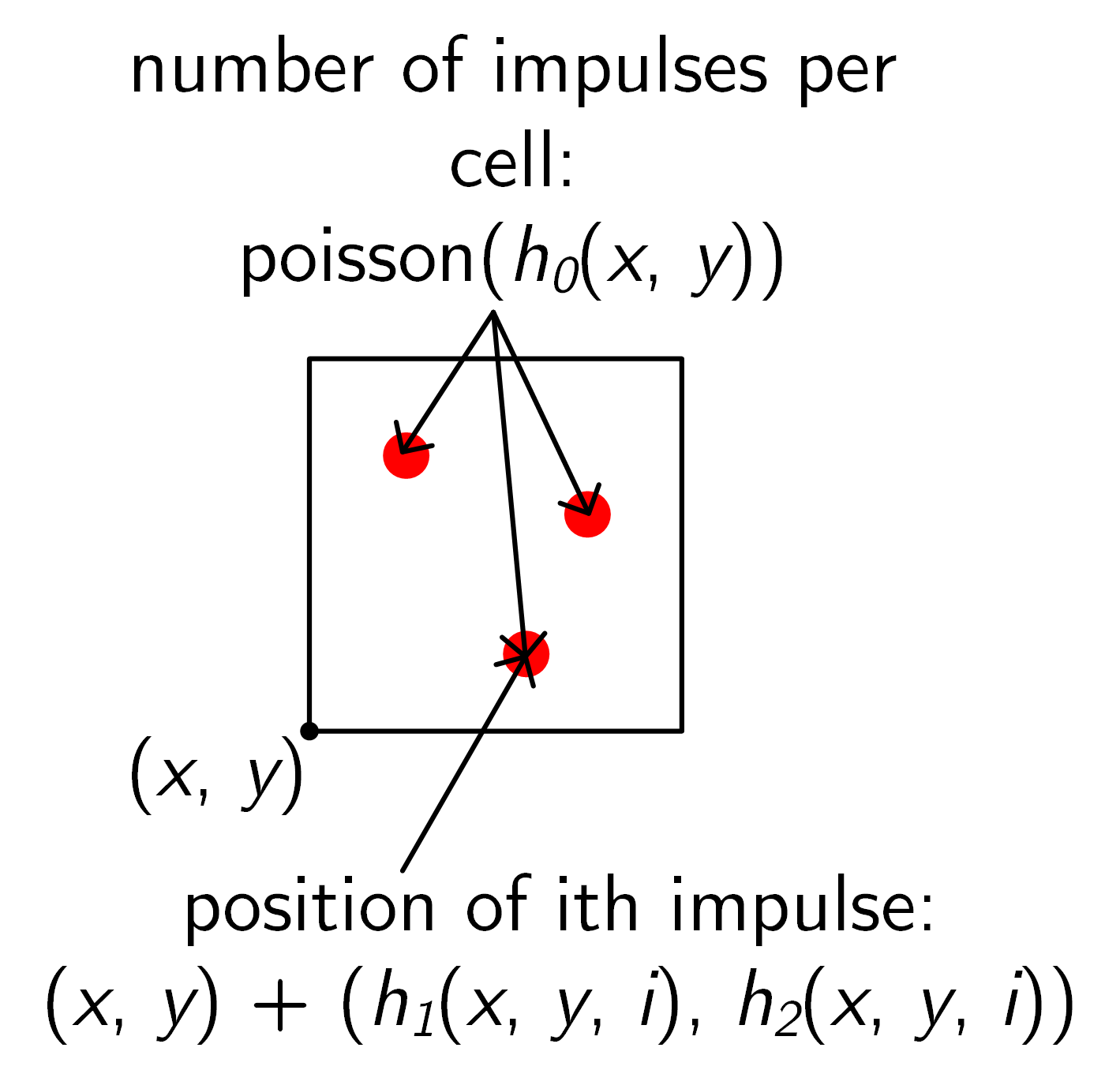} \\
(c) \includegraphics[width=0.4\linewidth]{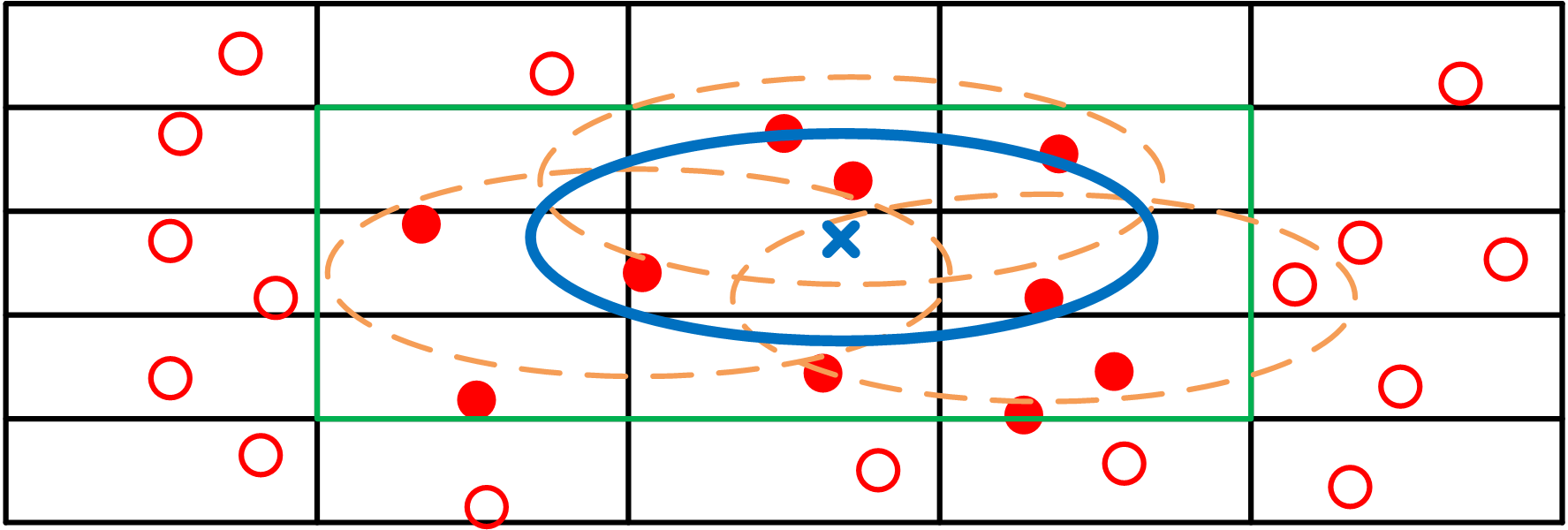} &
(d) \includegraphics[width=0.4\linewidth]{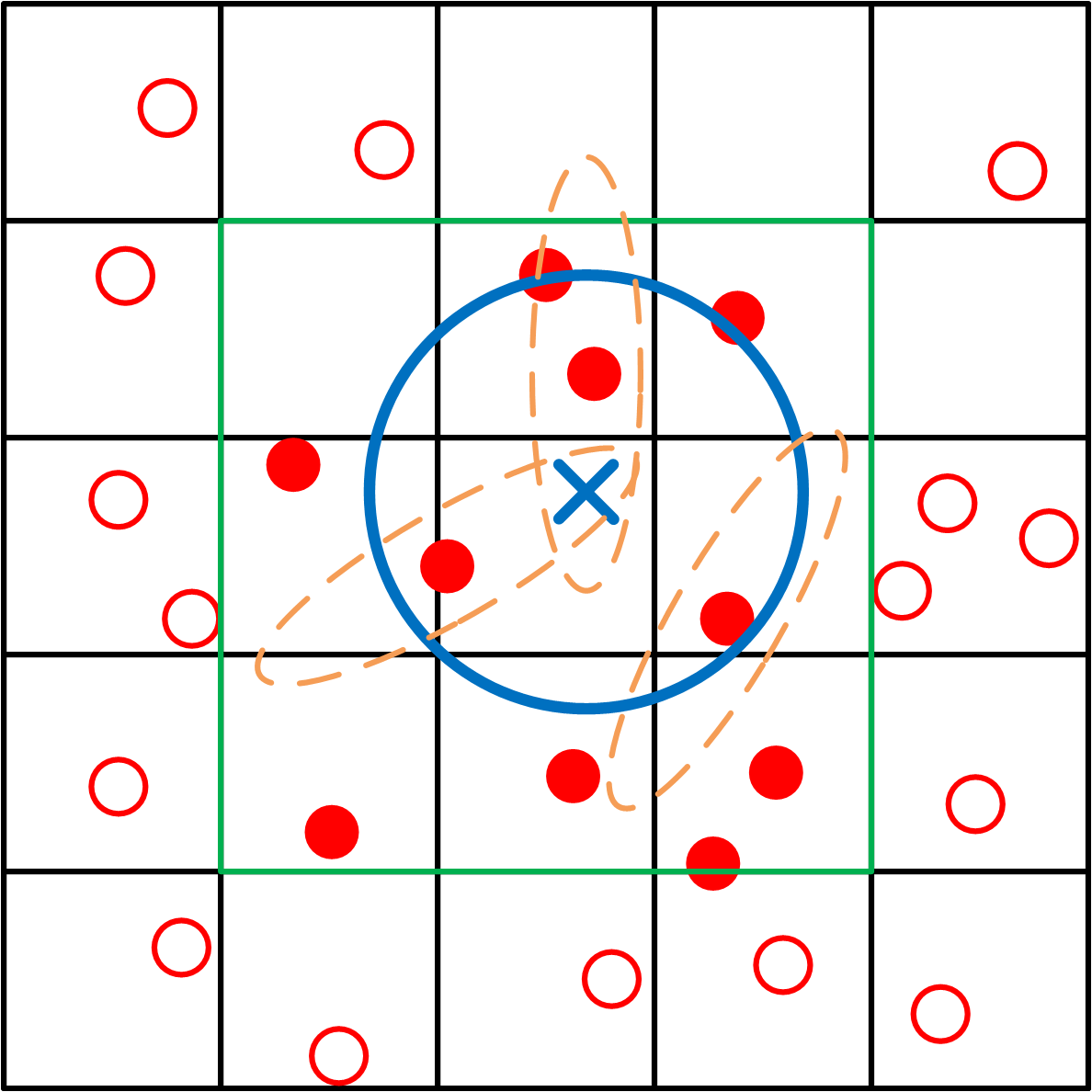} \\
(e) \includegraphics[width=0.4\linewidth]{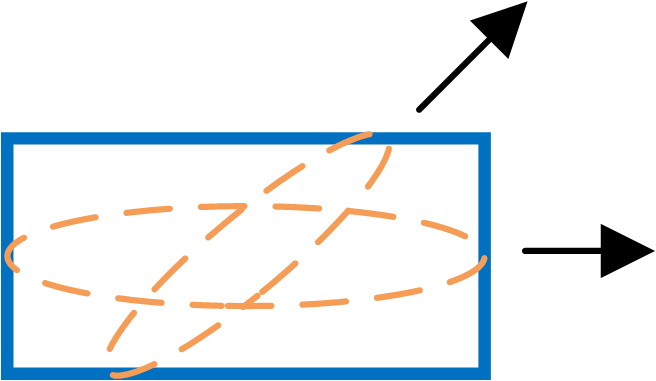} &
(f) \includegraphics[width=0.4\linewidth]{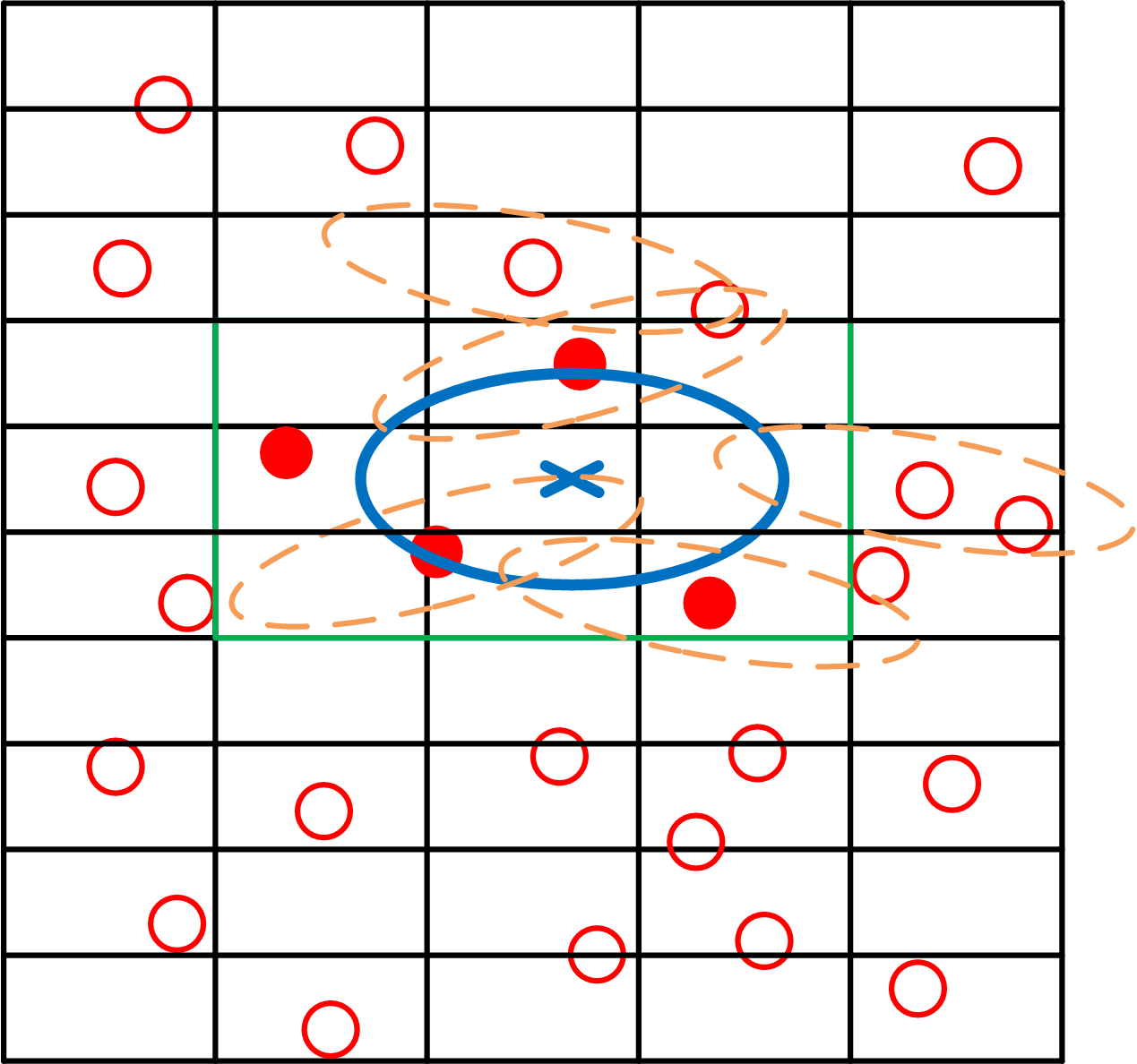}
\end{tabular}
\caption{\textbf{(a)} To find nearby impulses we determine the grid cell that a query point falls in. Then we determine the number and position of impulses in that cell and neighboring cells. \textbf{(b)} To determine the number of impulses in a grid cell we hash the cell indexes into a Poisson distribution. Then for each impulse we hash the cell indexes along with the impulse index again to determine its position within the cell. \textbf{(c)} For elongated axis-oriented kernels we can stretch the cells. This does not decrease the efficiency (the ratio of the number of impulses affecting the query point to the number of impulses considered). \textbf{(d)} For arbitrarily oriented kernels we must conservatively bound the volume they may affect with a sphere. This results in a large grid cell volume relative to the kernel volume and decreases the efficiency. \textbf{(e)} If the kernels are approximately axis-aligned we can decrease the width of the grid cells in the non-favored directions and rescale the non-aligned impulses to fit. \textbf{(f)} This ``hybrid'' scheme allows a tradeoff between the efficiency of (c) and the orientability of (d).}
\label{fig:impulse}
\end{figure}

To do this we use the usual (e.g. \cite{Lagae:2009:SparseGabor}) method of dividing $\mathbb{R}^3$ into a regular grid of cells. For a given query point and kernel radius, we can find nearby impulses by checking nearby cells. We can generate a white noise (uniform) distribution by assigning a number of impulses to each cell according to a Poisson distribution, then positioning each such impulse uniformly at random within the cell. To ensure consistency between queries, we can generate the number of impulses in each cell by hashing the grid cell index, then generate the position of each impulse within each cell by hashing the cell index and impulse index. This process is shown in Figure \ref{fig:impulse}.

\begin{figure}
\centering
\includegraphics[width=0.4\linewidth]{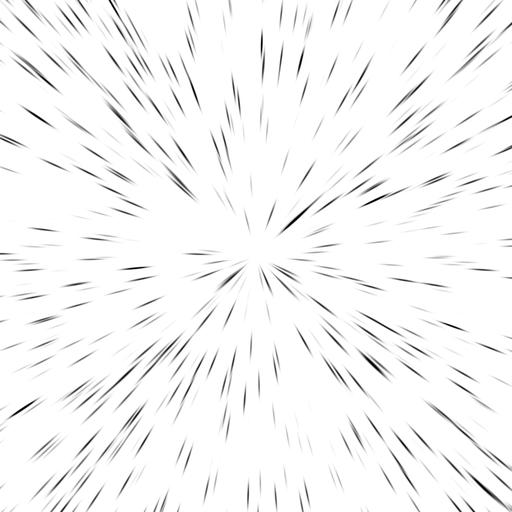}
\includegraphics[width=0.4\linewidth]{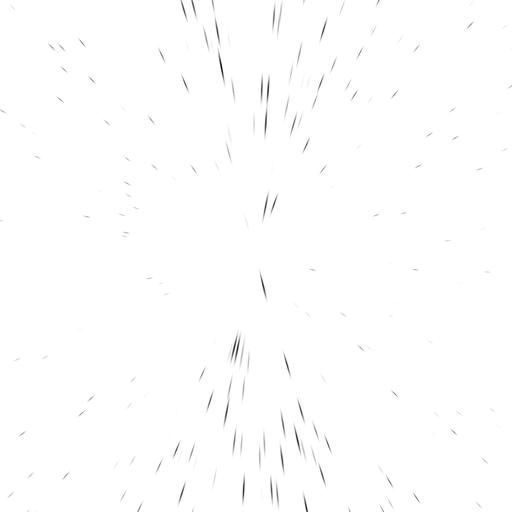}
\caption{\textbf{Left:} The simple scheme allows us to orient these prolate spheroids arbitrarily. However, since we must size the cells according to the major axis of the prolate spheroids, this comes at a high computation cost. \textbf{Right:} If the underlying vector field is known to be approximately axis-aligned, we can shrink the cells in the non-aligned dimensions. This results in fewer impulses that need to be considered while still delivering good results in the favored direction; this example demonstrates spheroid kernels in all directions with the vertical direction being favored. In three dimensions this example results in a factor of 16 fewer impulses that need be considered.}
\label{fig:oriented}
\end{figure}

\subsection{Non-Cartesian cells}
\begin{figure}
\includegraphics[width=\linewidth]{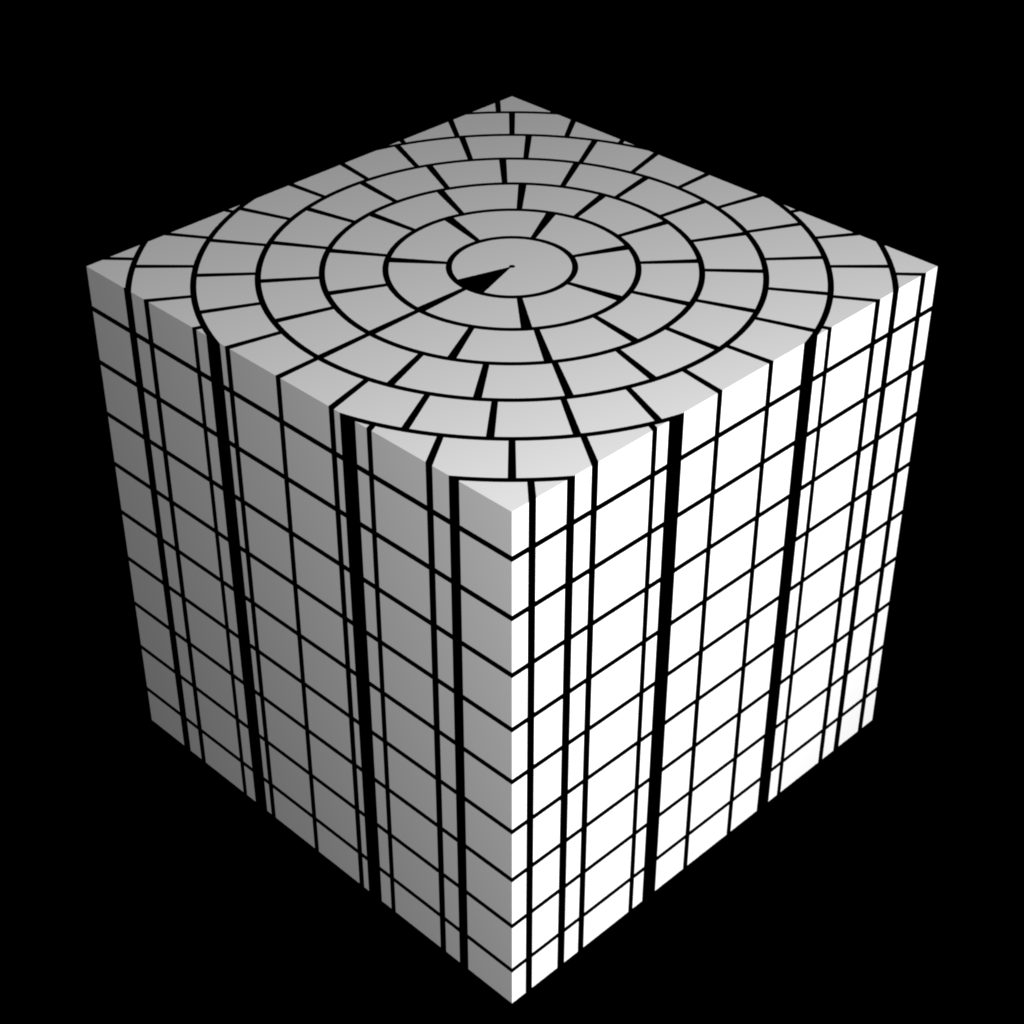}
\caption{Cells appropriate for cylindrically-aligned kernels.}
\label{fig:cylindrical-grid}
\end{figure}

For efficient local search it is essential that the cells have similar shape and size to the kernels. While some features, such as pores, are approximately aligned with a Cartesian axis, others, such as rays, are not. Therefore, when appropriate we use a ``cylindrical'' cell scheme. As depicted in Figure \ref{fig:cylindrical-grid}, the cells are arranged in concentric cylindrical bands where the number of cells in each band is proportional to the midpoint radius of the band, thus ensuring that each cell has the same volume. The neighborhood of a point can then be determined by calculating which cell it would have fallen into if it had landed in each of the adjacent bands.

\subsection{Kernels}

\begin{figure*}
	\begin{center}
		\includegraphics[width=0.24\linewidth]{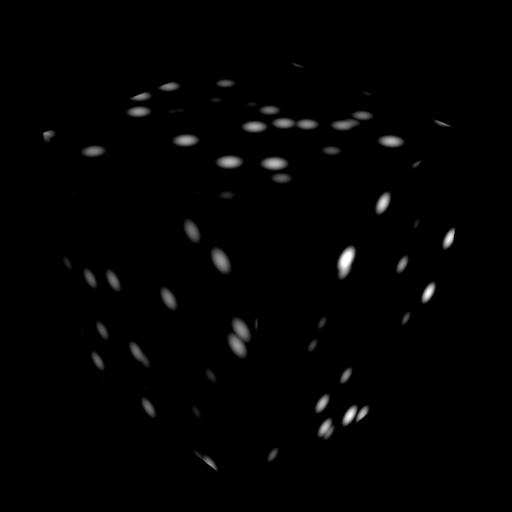}
		\includegraphics[width=0.24\linewidth]{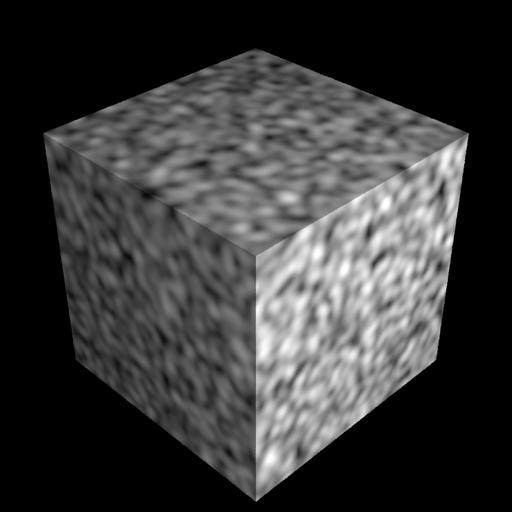}
		\includegraphics[width=0.24\linewidth]{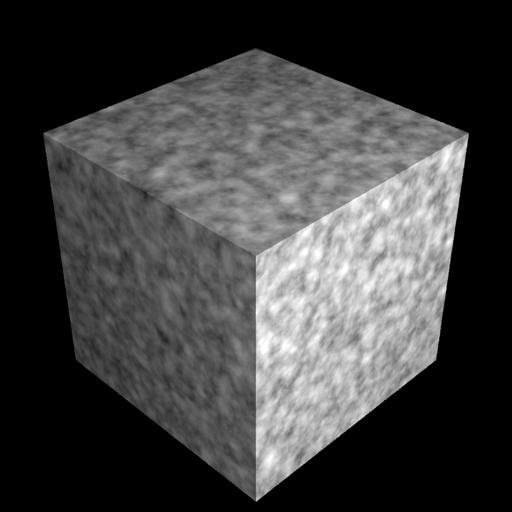}
		\includegraphics[width=0.24\linewidth]{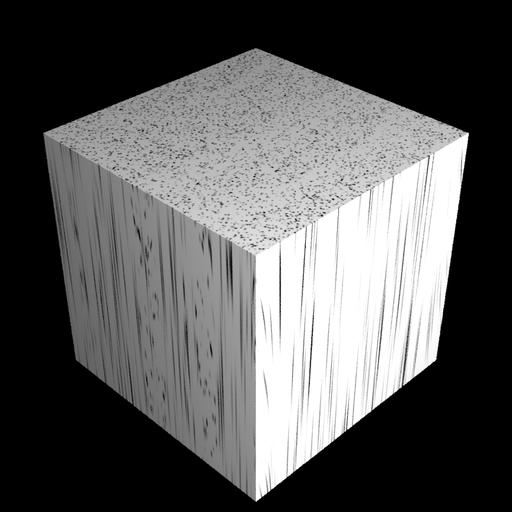}
	\end{center}
	\caption{Sparse convolution noise. Low impulse densities (far left) are appropriate for discrete features, while high impulse densities (center left) create noise. Multiple bands (center right) can be added together to create features at a spread of sizes. Oriented noise (far right) helps capture effects such as interlocked grain.}
	\label{fig:noise}
\end{figure*}

For ``continuous'' noise we use the Wyvill kernel as described in \cite{Shirley:2009:Fundamentals}:
\begin{align}
	K \left( \mathbf{p} - \mathbf{k} \right) = K\left(r\right) = \left(1 - r^2\right)^3 && r \in \left[0, 1 \right)
\end{align}
where $r$ is the distance from the point $\mathbf{p}$ to the kernel center $\mathbf{c}$:

\begin{align}
	r = \left\lVert \mathbf{p} - \mathbf{k} \right\rVert
\end{align}

The compact support of the Wyvill envelope allows us to avoid the tradeoff between a wider search radius and the small but sharp step resulting from truncation. Though it is perhaps not as analytically elegant as a non-truncated Gabor kernel (\cite{Lagae:2009:SparseGabor}), we find that it produces good results in practice.

For discrete features with sharper borders we instead use a variation on the canonical bump kernel
\begin{align}
	K\left(r\right) = e^{-sr^2 / \left(1 - r^2\right)} && r \in \left[0, 1 \right)
\end{align}
where $s \geq 0$ is a "sharpness" parameter that blends between a box function for $s \rightarrow 0$, a bump for $s \approx 1$, and a spike for $s \rightarrow \infty$.

\subsection{Multiband and anisotropic noise} 

For less-structured patterns, we sum multiple bands of noise. To simplify the design of multiband noise, we have elected to begin with a base band and then generate additional bands with geometrically smaller radii based on a power law with a typical magnitude dropoff with kernel size with exponent $\gamma$ between 1 and 2.

\begin{figure*}
	\begin{tabular}{cc}
		\begin{tabular}{c}
			\includegraphics[width=0.3\linewidth]{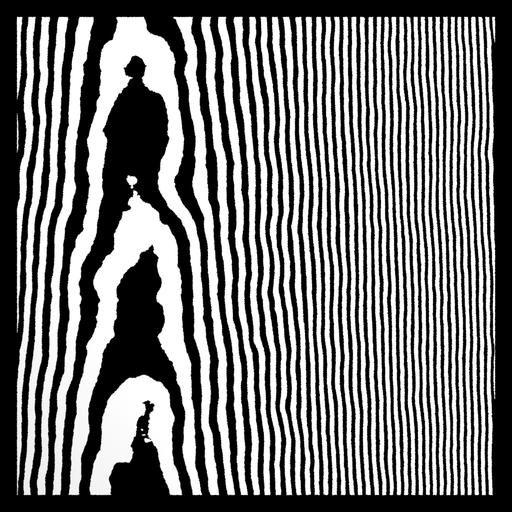} \\
			isotropic
		\end{tabular} &
		\begin{tabular}{rccc}
			\raisebox{0.07\linewidth}{fast} &
			\includegraphics[width=0.15\linewidth]{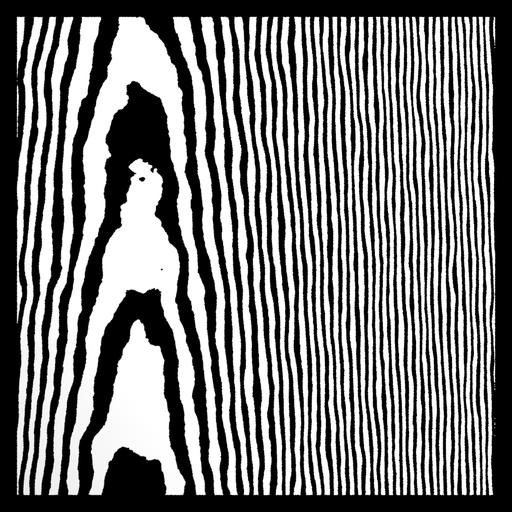} &
			\includegraphics[width=0.15\linewidth]{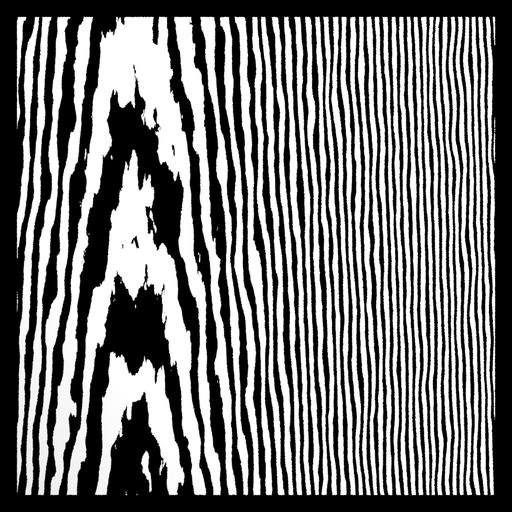} &
			\includegraphics[width=0.15\linewidth]{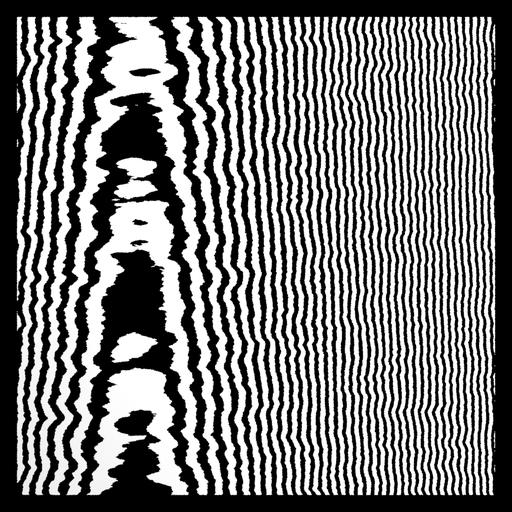} \\
			\raisebox{0.07\linewidth}{slow} &
			\includegraphics[width=0.15\linewidth]{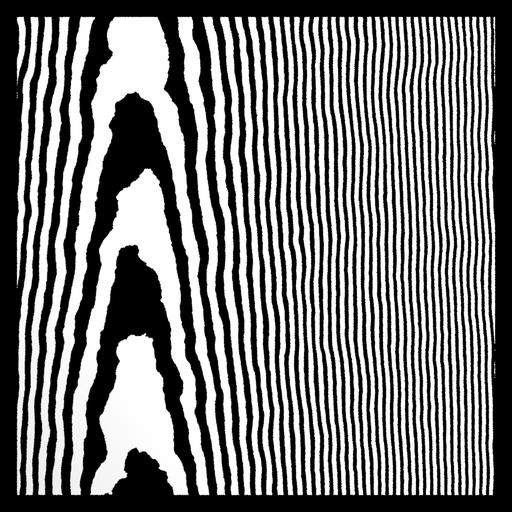} &
			\includegraphics[width=0.15\linewidth]{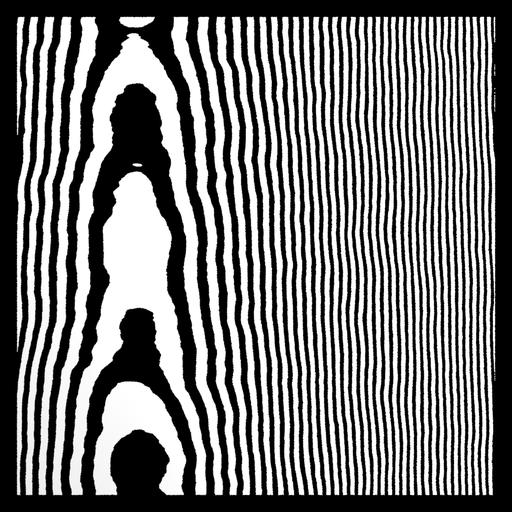} &
			\includegraphics[width=0.15\linewidth]{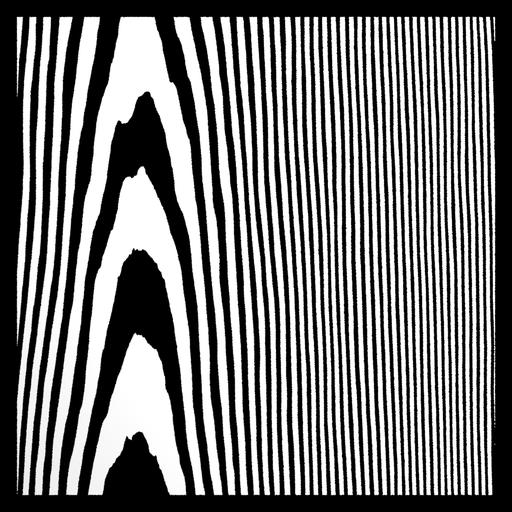} \\
			& $r$ & $\theta$ & $z$
		\end{tabular}
	\end{tabular}
	\caption{Effect of varying the distortion envelope size on ring patterns.}
	\label{fig:envelope}
\end{figure*}

Furthermore, we allow the kernels to be stretched into cylindrical-axis-aligned ellipsoids. When applied as a distortion in the radial direction, the width of the kernel in each of the three directions corresponds to a distinctive change in the growth ring pattern as shown in Figure \ref{fig:envelope}.

Symbolically, we may define the base band as
\begin{align}
	N_0 \left( \mathbf{p} \right) = \alpha \sum_k K \left( S^{-1} \left( \mathbf{p} - \mathbf{k} \right) \right)
\end{align}
where $\alpha_0$ is a scaling of the magnitude of the base band, $k$ are the kernel centers, and $S$ is a scale matrix corresponding to the size of the cells/kernels.

For multiband noise, the higher frequency bands are then
\begin{align}
	N_i \left( \mathbf{p} \right) = \alpha \beta^{\gamma i} \sum_k \beta^i K \left( S^{-1} \left( \mathbf{p} - \mathbf{k} \right) \right)
	\label{eq:multiband_isotropic}
\end{align}
where $\beta$ is the size factor from one band to the next, typically near $\frac{1}{2}$.

\subsection{Oriented noise}

It is useful to be able to orient these non-isotropic kernels using an underlying frame field, which can be defined using a rotation matrix $R$ at each point in space. \cite{Lagae:2009:SparseGabor} achieves oriented noise by directing a harmonic component while keeping the envelope isotropic.

We have chosen to scale the entire kernel instead of using a harmonic component. Therefore, to orient noise, we instead align the entire kernel with the frame field. However, this simple approach is inefficient for envelopes with extreme aspect ratios, since for determining the cell neighborhood we need to conservatively bound the volume that the envelope covers with a sphere of radius equal to the longest axis of the envelope. For prolate spheroids, the efficiency (ratio of kernels affecting a particular point versus kernels considered) decreases as the square of the ratio of the axes.

Fortunately, in our cases the features are all \textit{approximately} axis-aligned, either in the radial or longitudinal directions. We therefore decouple the cell scaling $S_c$ from the kernel scaling $S_e$.

For the cells we use aspect ratios partway between unity (as in the simple approach) and the aspect ratio of the kernels. Naturally the cell size must still remain fixed across space.

As for the kernels, we allow their scales, aspect ratio, and orientation to vary across space. To ensure the kernels do not extend beyond the search radius, we apply a uniform scaling to each kernel so that it fits within one axis aligned-cell centered on the kernel. This can be done by finding the axis-aligned bounding box of the rotated and scaled ellipsoid. \cite{Barnes:2014:ExactBoundingBoxes} shows that if a sphere is scaled into an ellipsoid by $S_e$ and then rotated by $R$,
the axis-aligned bounds of the result is determined by the 2-norms of the rows of $R S_e$:

\begin{align}
	x_\text{max} = \sqrt{\left( RS_e \right)_{xx}^2 + \left( RS_e \right)_{xy}^2 + \left( RS_e \right)_{xz}^2 } \\
	x_\text{min} = -x_\text{max}
\end{align}

and likewise for $y$ and $z$. The ratio of these to the corresponding cell dimension then determines the necessary scaling:

\begin{align}
	s_b = \min \frac{S_{cx}}{x_\text{max}}, \frac{S_{cy}}{y_\text{max}}, \frac{S_{cz}}{z_\text{max}}
\end{align}

Equation \ref{eq:multiband_isotropic} then becomes

\begin{align}
	N_i \left( \mathbf{p} \right) = \alpha \beta^{\gamma i} \sum_k \beta^i K \left( s_b^{-1} S_e^{-1} R^T \left( \mathbf{p} - \mathbf{k} \right) \right)
\end{align}

A two-dimensional depiction of this scheme is shown in Figure \ref{fig:impulse} and example results in Figure \ref{fig:oriented}.

We query $R$ and $S_e$ at the evaluation point $p$ rather than the kernel center $k$, as this requires fewer queries in a random-access scheme.

\subsection{Gradients}

In addition to the value of the kernel, we will also need the gradient. This is

\begin{align}
	\nabla K \left( s_b^{-1} S_e^{-1} R^T \left( \mathbf{x} \right) \right)
	=& R S_e s_b \nabla K \left( s_b^{-1} S_e^{-1} R^T \left( \mathbf{x} \right) \right) \\
	& \left(+\; O \left( \nabla S_e, \nabla R \right) \right)
\end{align}

In practice we omit the trailing terms in $\nabla S_e$ and $\nabla R$, as we expect $S_e$ and $R$ to vary little across the extent of the kernel.

\section{Distortion} \label{sec:distortion}

Noise is useful not only for scalar values (e.g. generating pore masks), but also for spatial structure: while we can imagine that idealized wood grows in perfect cylinders, real wood departs from this, both in terms of distinctive figure and less distinct random variations. To model these, we begin with an idealized model of wood, then distort the lookups into the model using a point field $\mathbf{f} : \mathbb{R}^3 \rightarrow \mathbb{R}^3$.

\subsection{Decomposition into direction and magnitude}

To simplify the development and analysis of distortion, we decompose the point field into two parts: a normalized vector-valued direction field $\mathbf{a}$ and a scalar-valued magnitude field $m$. We then define $\mathbf{f}$ as 
\begin{align}
	\mathbf{f} \left( \mathbf{p} \right) = \mathbf{p} + m \left( \mathbf{p} \right) \mathbf{a}\left( \mathbf{p} \right)
\end{align}

As with any other scalar-valued function, $m$ may be generated using a noise function as described previously. Imperfections in the noise function seem to be much less visually obvious in distortions compared to a direct color visualization of the noise, so we can get away with a lower impulse density here. Even a mean of just two kernels per band overlapping each point is usually sufficient to produce satisfactory results.

$\mathbf{a}$ may be as simple as a coordinate direction, e.g. $\mathbf{\hat{r}}$. This allows us to design distortions that correspond qualitatively to the types of growth ring and fiber direction variation seen in real wood. A distortion in the $\mathbf{\hat{r}}$ direction affects the shape of the growth rings and produces what \cite{Beals:1977:FigureInWood} calls blister or quilted figure, whereas a distortion in the $\mathbf{\hat{\theta}}$ direction results in curly or fiddleback figure. See Figures \ref{fig:distortion_radial} and \ref{fig:distortion_tangential} for examples.

With these distortions in hand, we can replace the simple lookup $s \left( \mathbf{p} \right)$ of a scalar-like value, such as color, with 
\begin{align}
	s \left( \mathbf{f} \left( \mathbf{p} \right) \right)
\end{align}

\subsection{Transformation of directional values}

In addition to scalars, we may also need to distort directional values, such as vector and gradient fields. This is particularly important for wood since the anisotropic reflectance behavior depends strongly on the local fiber direction. 



However, merely modifying the lookup does not produce the desired result for vectors and gradients, since these values themselves are also spatial in nature. For example, merely changing the lookups into a longitudinal fiber direction field does not result in any change since the fiber direction is the same everywhere. This would make it impossible to produce the figure shown in Figure \ref{fig:distortion_radial}. Modifying the fiber direction independently of the distortion applied to scalar values does not produce the distinctive correlation between the growth ring pattern and the figure.

Rather, we need to adjust the result using the Jacobian of the distortion. Vectors $\mathbf{v}$ vary contravariantly, while gradients $\mathbf{g} = \nabla s$ vary covariantly:
\begin{align}
	J_\mathbf{f}^{-1} \mathbf{v} \left( \mathbf{f} \left( \mathbf{p} \right) \right) & \\
	J_\mathbf{f}^T \mathbf{g} \left( \mathbf{f} \left( \mathbf{p} \right) \right) & = \nabla s \left( \mathbf{f} \left( \mathbf{p} \right) \right)
\end{align}
Note that we use the identity $J_\mathbf{f^{-1}} = J_\mathbf{f}^{-1}$. This means we do not need to know what the inverse of $\mathbf{f}$ is, or even guarantee that it exists.

The Jacobian of the simple distortion function described earlier is
\begin{align}
	J_\mathbf{f} =& I + \mathbf{a} \otimes \nabla m \label{eq:jacobian} \\
	& \left(+\; m J_\mathbf{a} \right)
\end{align}
where $\otimes$ denotes an outer (tensor) product. For a constant vector field the last, parenthesized term is identically zero. Even when it is not, it is usually negligible in typical use cases and may be omitted with no obvious visual impact.

%

\subsection{Dealing with foldover}

Examining Equation \ref{eq:jacobian}, the volume texture may fold over (i.e. map a simple curve in the lookup to a self-intersecting curve in the source volume) if the condition
\begin{align}
	\left\| \nabla m \right\| < 1 \label{eq:gradient_condition}
\end{align}
fails to be met. Unfortunately, the gradient of sparse convolution noise is not bounded for any positive uniform impulse density, even if its magnitude is much smaller than 1 over most of the volume. In general it is inconvenient to restrict $m$ to the set of functions which satisfies (\ref{eq:gradient_condition}) everywhere.

The obvious potential problem of \text{not} making such a restriction is that parts of the volume texture may be repeated. In practice this turns out not to be visually detectable. The distortion occurs in three dimensions, but the viewer typically only sees a two dimensional cut through the surface. Therefore, even if a particular visible point is duplicated, it is unlikely that its double also lies on the visible surface. It is even less likely for this to hold for a large, contiguous, and undistorted enough set of visible points to produce a recognizable duplicated patch. Finally, the anisotropic reflectance of wood can only serve to further obscure such duplication.

A mathematically subtler but more visually obvious problem is that if the condition (Equation \ref{eq:gradient_condition}) is not met, the Jacobian $J_\mathbf{f}$ may pass through a singularity, causing a discontinuity in distorted vector-like values. To solve this, we simply ``cheat'' in the computation of the Jacobian by compressing the gradient to lie within a unit sphere. Omitting the parenthesized term, we modify Equation \ref{eq:jacobian} to

\begin{align}
	J_\mathbf{f} \approx I + \frac{\mathbf{a} \otimes \nabla m}{1 + \left\| \nabla m \right\|}
\end{align}

This ensures that $J_\mathbf{f}$ is positive-definite everywhere and thus never passes through a singularity while not having much effect on small gradients.

\section{Shade evaluation structure}

With these basic tools in hand, we can go about assembling them into a complete procedural wood shader. Our shade evaluation strategy is much like Shade Trees \cite{Cook:1984:ShadeTrees}. Our shader is organized as a  directed acyclic graph of nodes. Given a query point, the sink node recursively queries other volume texture nodes until a set of source nodes is reached. The results are then propagated back to the sink node. Individually, the nodes are fairly simple, but they can be combined to simulate the complex effects found in wood.

A diagram of the graph structure for our wood model is shown in Figures \ref{fig:pre_distortion} and \ref{fig:post_distortion} and a detailed description follows. Note that we describe the model progressing from the sources to the sinks here, but in practice a renderer intersection is produced at the sink, queries are propagated \textit{backwards} to the sources, and the results are returned forwards to the sink.

\subsection{Pre-distortion}


\subsubsection{Cylindrical coordinates}

The starting point for our volume texture is cylindrical coordinates. This defines three scalar fields corresponding to the $r$, $\theta$, and $z$ values. The $r$ value forms the basis for representing the seasonal growth of the tree.

The gradients of the coordinates produce three orthogonal vector fields, or equivalently, a frame field. This is useful for aligning fibers: the main longitudinal fibers are aligned in the $z$ direction, and the radial fibers of rays in the $r$ direction. Likewise, discrete features such as rays and pores are aligned using this frame field.

\subsubsection{Noise}

Noise is used several times in the volume texture DAG, both in 1D and 3D variants and in single and multiple bands. The parameters of the noise are as follows:

\begin{itemize}
	\item A kernel, in our case either the Wyvill kernel or the bump kernel.
	\item The envelope size, which is an axis-aligned scaling applied to the unit sphere (a $d$-tuple for $d$ dimensions).
	\item The impulse density, i.e. the mean number of kernels overlapping any point in space.
	\item The magnitude of the kernel values.
	\item For multiband noise, a number of bands, the frequency factor between successive bands (usually about 2), and the power dropoff with frequency (usually around $f^{-1}$).
\end{itemize}

The ellipsoidal kernels used to generate the noise may be axis-aligned, or they may be aligned using a frame field.

\subsubsection{Seasonal growth}

We turn the $r$ coordinate volume into a time volume by scaling its value and adding a smoothed triangular wave $w\left(r\right)$ to it. Each cycle of the smoothed triangle wave consists of two linear segments connected by quadratic segments so as to make the wave $C^1$ continuous. The parameters are the lengths of the two linear segments and the two quadratic segments.

We scale the triangle wave so that when added to the radius, half of the time the wood grows faster than the mean and half of the time the wood grows more slowly. Then we add 1D noise $n$ to generate variation in growth rate. All in all, we have

\begin{align}
	t_\text{pre} &= r + w\left(r\right) \\
	t &= t_\text{pre} + n \left( t_\text{pre} \right)
\end{align}

This gives us the time at which any point in the wood was laid down.

\subsubsection{Growth rings}

We can then create growth rings from the time volume. By using a smoothed rectangular wave $g = w\left(t\right)$, we can vary the absorption of the wood according to the season. Similarly to the triangle wave used to determine the growth rate, each cycle of the smoothed rectangular wave consists of two constant segments and two cubic segements so as to make the wave $C^2$ continuous, and the parameters are the relative lengths of the two linear segments and the two quadratic segments, plus the minimum and maximum values.

\subsubsection{Interlocked grain}

At the same time, we take the cylindrical frame field and rotate it about the $r$ axis at each point. The angle of rotation is determined by 1D noise based on the $r$ coordinate. This produces a frame field that simulates the patterns of fiber directions found in interlocked grain.

\begin{figure}
	\begin{tabular}{rr}
		(a) \includegraphics[width=0.4\linewidth]{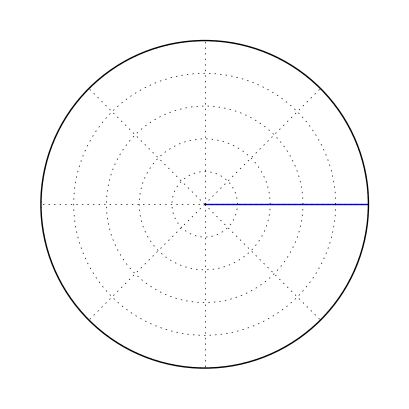} &
		(b) \includegraphics[width=0.4\linewidth]{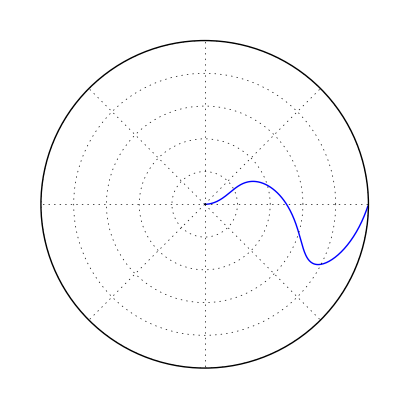} \\
		(c) \includegraphics[width=0.4\linewidth]{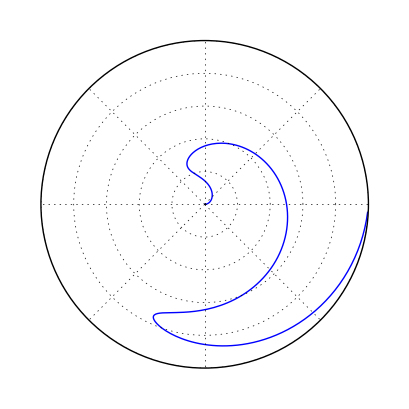} &
		(d) \includegraphics[width=0.4\linewidth]{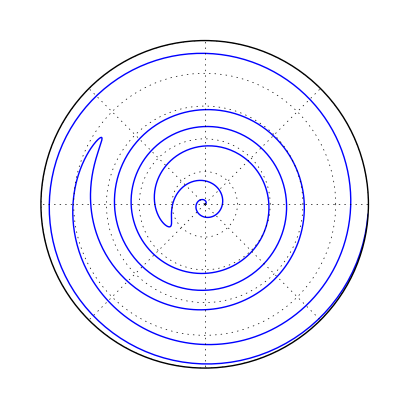} 
	\end{tabular}
	\caption{Problems with attempting to represent interlocked grain via distortion. \textbf{(a)} Consider a radial slice in the undistorted wood. \textbf{(b)} To represent the helical pattern of fibers, a distortion would have to displace points circumferentially ($\theta$) proportionally to the longitudinal ($z$) position of the point. The direction of the helix changes with radius. As we move in the $z$ direction the distortion turns a radial line into a wave. \textbf{(c)} As we continue to move in the $z$ direction, the waviness becomes more severe. \textbf{(d)} The required displacement (in terms of $\theta$) increases without bound. Our solution is to use distortion only for local variations, and align features with a vector field representing the interlocked grain at the time they are originally generated.}
	\label{fig:interlocked}
\end{figure}

Note that interlocked grain cannot be adequately represented by simply distorting a cylindrically-aligned frame field. The problems with attempting to do so are illustrated in Figure \ref{fig:interlocked}.

\subsubsection{Discrete features}

With the frame field and the growth ring volume in place, we then generate discrete features using a noise function with low impulse density and the bump kernel with tunable sharpness.

Although in real wood rays do not normally end and pores generally run the length of the tree, it is convenient to approximate them as elongated but finitely long kernels in order to easily accommodate changes in the frame field over large distances. In order to avoid evaluating the frame field more than once per query, we use the frame field at the position of the query rather than at the kernel center.

To represent ring-porous woods, we can modulate the pore sizes using one minus the growth ring volume, which causes the pores to become vanishingly small in the latewood. By interpolating between this and the uniform pore size characteristic of diffuse-porous woods, we can represent varying degrees of ring-porousness. In the end, we produce a ray mask and a pore mask, representing the ``ray-ness'' and ``pore-ness'' of any point in space.

\subsubsection{Bump}

When negated and scaled by the pore radius, or some fraction thereof, the pore mask $p$ produces a height appropriate for a bump map.

\subsubsection{Color}

We offset and scale the growth ring volume $g$ and pore mask $p$ to produce a absorption path length volume. Then we use Beer's law and a color absorption parameter $\sigma$ to generate a color volume $c$. In all, we have:

\begin{align}
	c = e^{-\sigma \left(\ell_0 + \ell_g g + \ell_p p\right)}
\end{align}

To aid in tuning the model, we produce an estimate of $\sigma, \ell_0, \ell_g$ by taking the 25th and 75th percentiles of each channel in a single photograph and treating those as the earlywood and latewood colors. This produces a robust estimate at interactive speed.

\subsection{Distortion}

At this point we apply distortion as defined in \S \ref{sec:distortion} to all resulting volumes above---frame field, ray, pore, bump, and color. The distortion magnitude is based on multiband noise, with four bands usually being sufficient to produce pleasing results.

$r$ distortion is the most important as it is responsible for distinctive variation in the shape of growth rings when the wood is cut, as well as for many of the most dramatic types of figure. $\theta$ distortion does not change the shape of the growth rings, but can produce figure as well. $z$ distortion is the least important, as its only obvious effect is on rays. The parameters of the three distortions are controlled independently.


\subsection{Post-distortion}

\subsubsection{Wood BSDF}

After distortion we can begin assembling the wood BSDF directly. 

By extracting the longitudinal ($r$) and radial ($z$) directions from the frame field we have the fiber directions necessary for the fiber specular component of the wood BSDF for both the main fibers and the rays. The highlight width is given as a global parameter. The highlight color comes from the distorted color volume, which we also use for the diffuse component of the reflectance.

We use the ray mask to linearly interpolate between the BSDFs (\textit{not} the fiber directions) for these two fiber alignments. 

\subsubsection{BSDF composition}

At this point we have a complete wood BSDF. If desired, we may add auxiliary BSDFs such as the bump map described earlier, coatings, and other surface treatments. As with the bump map, any auxiliary BSDF can easily be made to determine its properties from the underlying wood volumes. For our renderings, we used a bump map based on the pore mask under a smooth or rough (Beckmann) coating.

\begin{figure}
\includegraphics{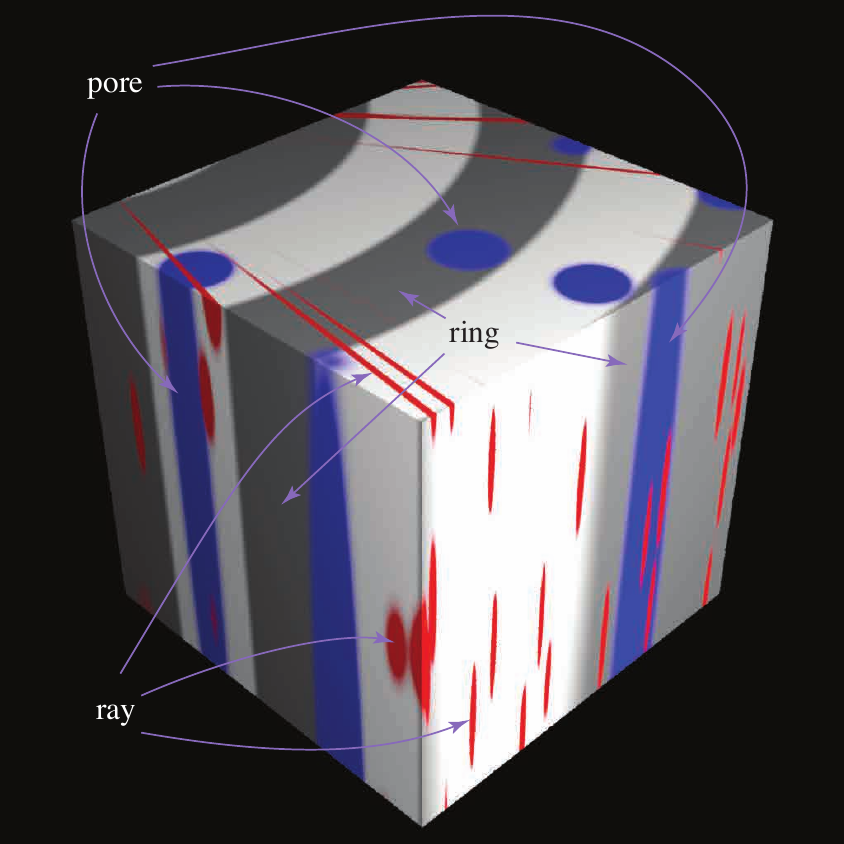}
\caption{A rendering of our model with no distortion and false colors to highlight the main component parts: rays, vessels, and growth rings.}
\end{figure}

\section{Results}

We implemented our model in the Mitsuba renderer \cite{Jakob:2014:Mitsuba}. We demonstrate our results on a variety of scenes. Still frames are shown here; a video is available at \url{https://www.youtube.com/watch?v=FG57kqn7GC8}.

\subsection{Overhead light sweeps}

\begin{figure}
	\begin{tabular}{lr}
		\includegraphics[angle=90, height=1.75in]{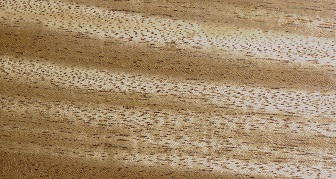} &
		\includegraphics[angle=90, height=1.75in]{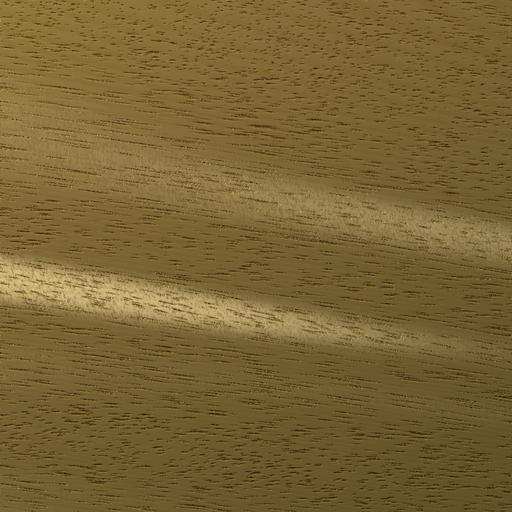} \\
		\includegraphics[angle=90, height=1.75in]{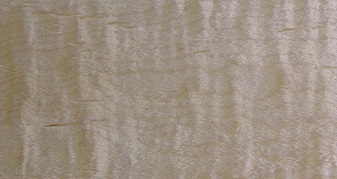} &
		\includegraphics[angle=90, height=1.75in]{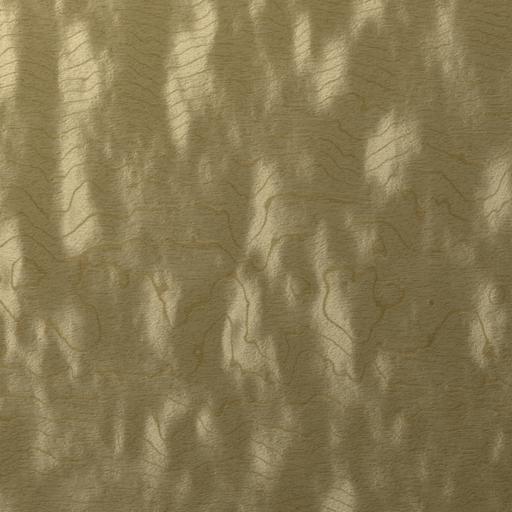} \\
		\includegraphics[angle=90, height=1.75in]{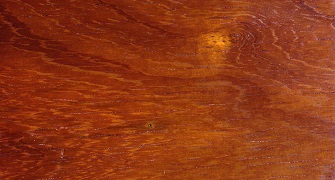} &
		\includegraphics[height=1.75in]{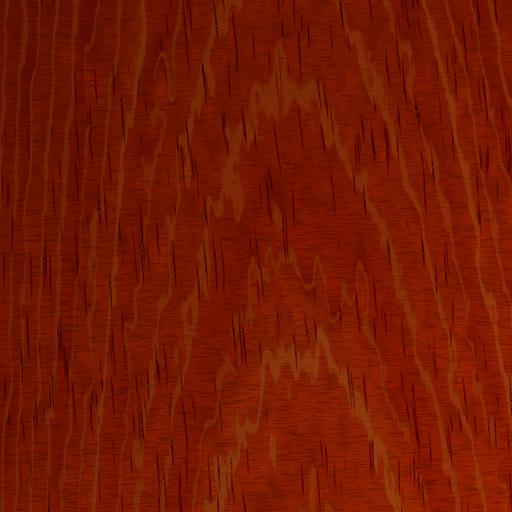} \\
		\includegraphics[angle=90, height=1.75in]{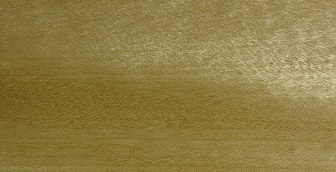} &
		\includegraphics[angle=90, height=1.75in]{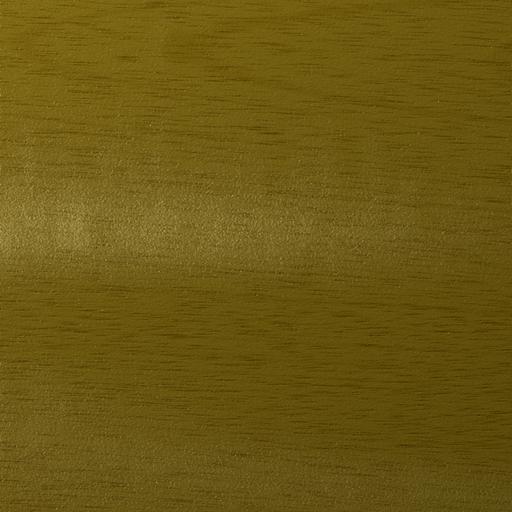}
	\end{tabular}
	\caption{Example still frames (left) and corresponding renderings (right) of procedural wood. From top to bottom: \textbf{African mahogany.} Note the highlight (stripe figure) due to interlocked grain and the corresponding pore shapes on the surface. \textbf{Tiger (curly) maple.} Radial variation in the growth produces wavy highlights known as curly figure. Note the correlation between the highlights and the growth rings. \textbf{Padauk.} Fast radial variation with respect to the circumferential direction produces these jagged growth rings. \textbf{Yellowheart.} Like the mahogany, this wood exhibits stripe figure due to interlocked grain.}
	\label{fig:sweeps}
\end{figure}

For the most direct comparison, we render blocks of wood from a fixed viewpoint as a light is swept overhead. Some still frames and corresponding renderings of procedural wood are shown in Figure \ref{fig:sweeps}.

\subsection{Wood floors}

\begin{figure}
	\includegraphics[width=\linewidth]{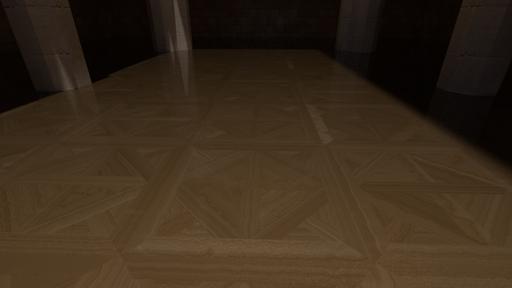}
	\caption{Nested-square parquetry floor pattern on the Sponza scene. Model by Marko Dabrovic, Kenzie Lamar, and Morgan McGuire; downloaded from Morgan McGuire's Computer Graphics Archive \url{http://graphics.cs.williams.edu/data}.}
	\label{fig:sponza}
\end{figure}

One common real-world application of wood is flooring, ranging from simple parallel board patterns to elaborate parquetry. A nested-square pattern is shown in the Sponza scene in Figure \ref{fig:sponza}. A board pattern can be defined by a function that takes a point in the surface space and produces:

\begin{itemize}
	\item A set of board indexes, which we can hash into some candidate region in the tree to determine where the origin of the board lies in the tree.
	\item An offset relative to the board origin, typically simply the difference vector to a designated corner of the board with a rotation to define the grain direction. Together with the above, this determines where in the tree the point came from.
\end{itemize}

\begin{figure}
	\includegraphics[width=\linewidth]{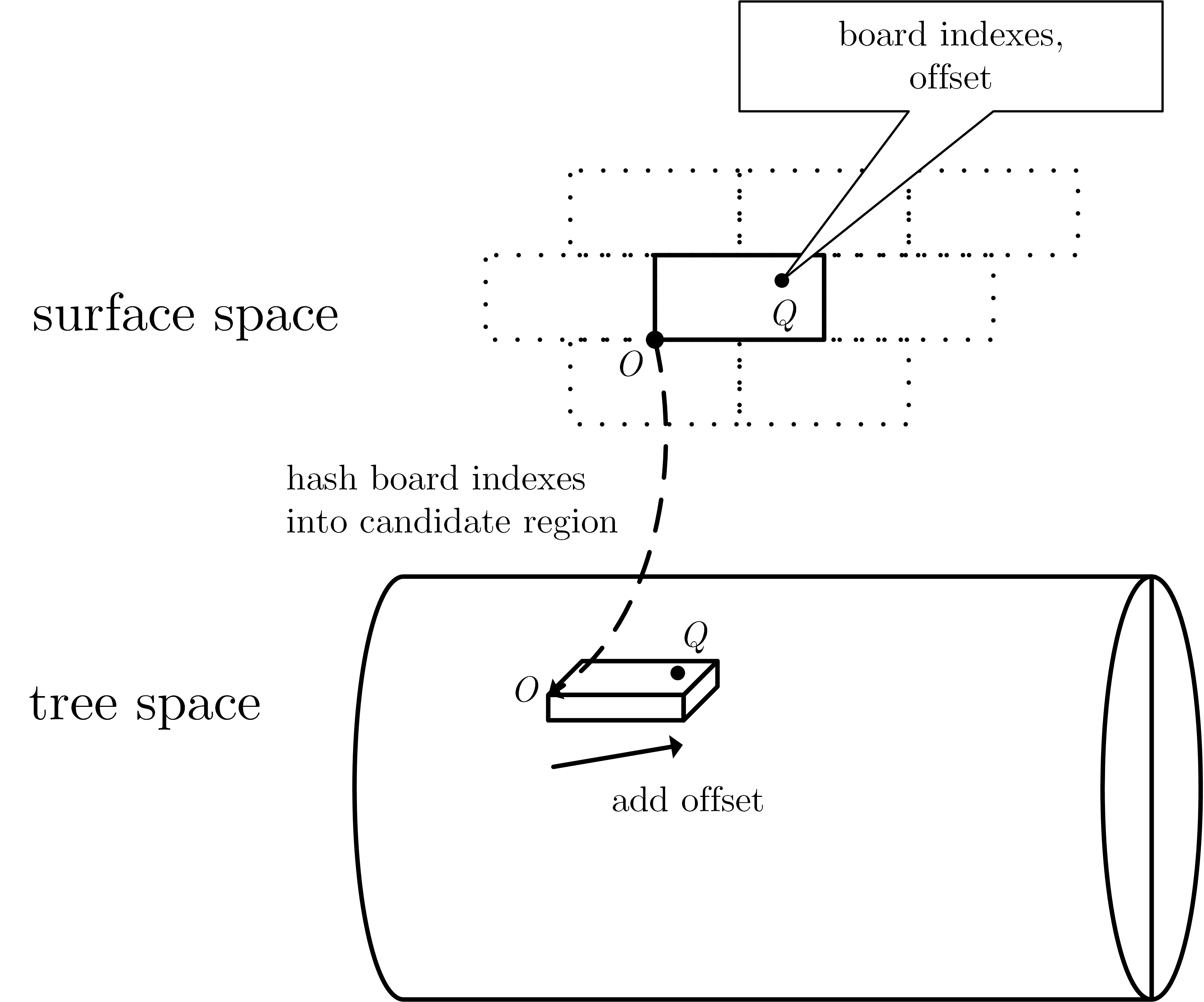}
	\caption{Operation of the ``board function''. Given a query point in surface space, the board function produces a set of board indexes and an offset. We hash the board indexes into some candidate region in the tree to find the origin of the board in the tree, then add the offset to produce the corresponding point in the tree.}
	\label{fig:board}
\end{figure}

This method is illustrated in Figure \ref{fig:board}.

\subsection{Objects}

\begin{figure}
	\includegraphics[width=\linewidth]{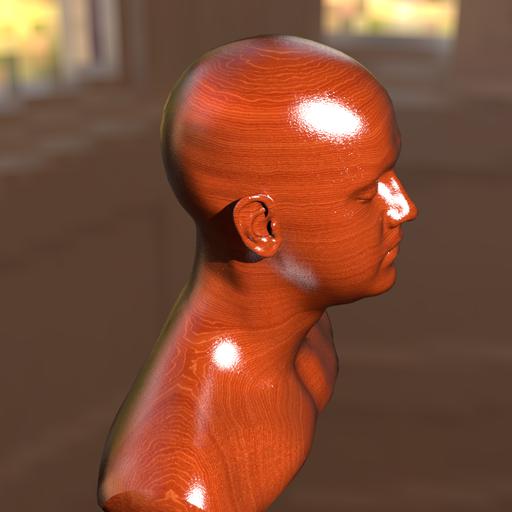}
	\caption{A carved head. Model by Lee Perry-Smith, Morgan McGuire, and Guedis Cardenas; downloaded from Morgan McGuire's Computer Graphics Archive \url{http://graphics.cs.williams.edu/data}. Environment map by Bernhard Vogl.}
	\label{fig:head}
\end{figure}

\begin{figure}
	\includegraphics[width=\linewidth]{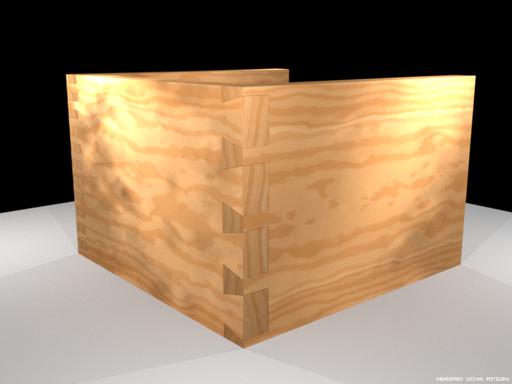}
	\caption{A partial box made out of multiple pieces.}
	\label{fig:joint}
\end{figure}

\begin{figure}
	\includegraphics[width=\linewidth]{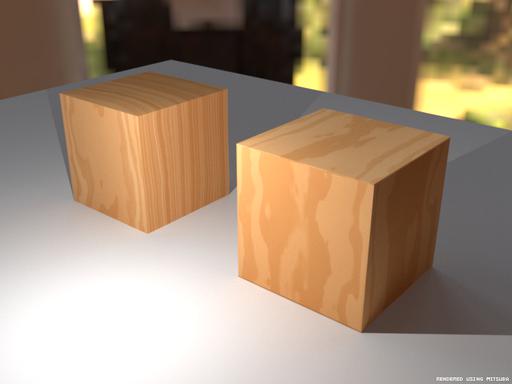}
	\caption{A comparison of a carved wooden cube (left) and a veenered one (right). The carved block shows one face of each of the three major cutting planes: tangential on the left, radial on the right, and transverse on the top. The surface of the veneered block is made of six distinct thin slices and shows a tangential face on all sides.}
	\label{fig:veneer}
\end{figure}

Objects may be carved out of a single piece of wood, made out of several wooden parts, or covered with wood veneer. Each has a distinctive appearance. A wooden head is shown in Figure \ref{fig:head}. A section of a box with one box joint and one dovetail joint is shown in Figure \ref{fig:joint}. A carved wooden cube is shown in \ref{fig:veneer}, along with a veneered counterpart. Note the difference in the grain pattern.

\subsection{Performance}

\begin{figure}
	\includegraphics[width=\linewidth]{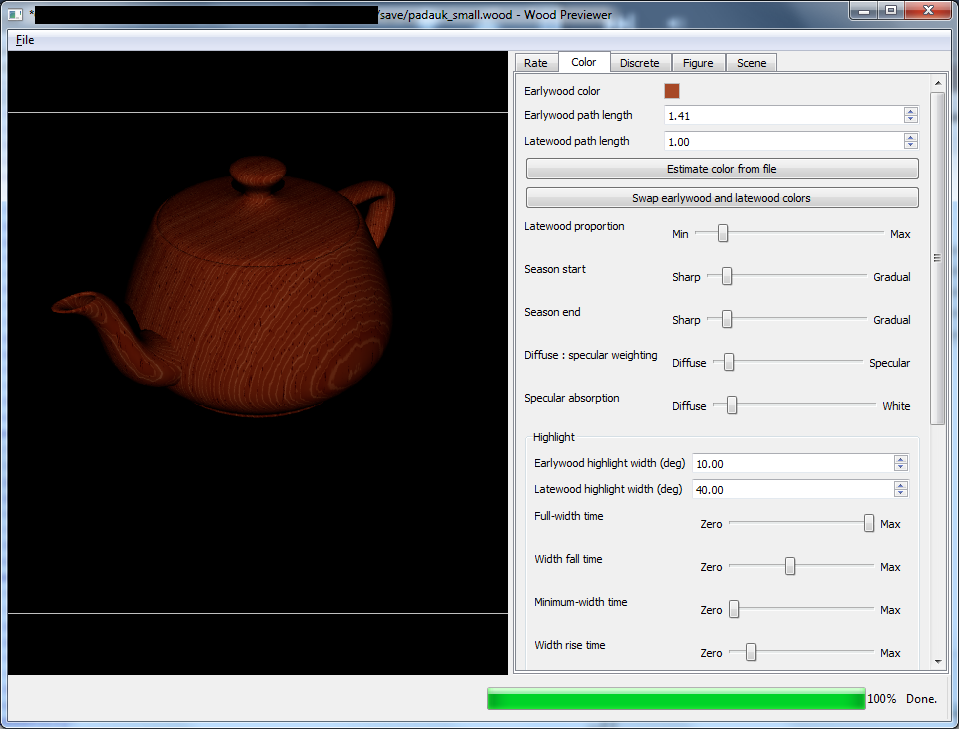}
	\caption{A screenshot of our wood previewer, which we used to generate the textures shown here. Model by Martin Newell, Jim Blinn, and Kenzie Lamar; downloaded from Morgan McGuire's Computer Graphics Archive \url{http://graphics.cs.williams.edu/data}. Environment map by Bernhard Vogl.}
	\label{fig:previewer}
\end{figure}

Our model can either be used inline at preview speed (see the video of our wood designer, a screenshot of which is shown in Figure \ref{fig:previewer}) or used to precompute traditional flat textures for quickly rendering the same object over many frames.

\section{Future work}

\subsection{Incorporating exemplar-based methods}
Although exemplar-based methods generally have the disadvantage that they do not \textit{allow} the user to tune the model other than by changing the exemplar, they do have the advantage that they do not \textit{require} the user to do so. To recover some of this advantage, we would like to come up with an automatic parameter estimation scheme for some or all of the parameters of our model, preferably requiring only a single photograph, which would give the user a quick starting point in the model that may then be adjusted as desired.

Furthermore, we could potentially improve the small-scale appearance of our model using exemplar-based techniques---effectively, a procedural model could generate a ``texture-by-numbers'' outline which would then be filled in by a exemplar-based subsystem similar to e.g. \cite{Ramanarayanan:2007:ConstrainedTextureSynthesis}.

\subsection{Knots} 
Knots are one prominent feature of wood that we have not addressed yet. They are the internal structure of branches embedded in the wood of the trunk.

\subsection{Revisiting the BSDF}

The wood BSDF model of Marschner \cite{Marschner:2005:FinishedWood} consists of independent diffuse and fiber specular components and assumes a constant chromaticity for each with respect to incident and outgoing angles. It may be worth deriving or approximating a BSDF using a more principled analysis of multiple scattering through the wood fibers, in a rough analogy to microfacet models versus the Phong model. The goal would be to produce more realistic results while reducing the number of parameters by unifying all color determination under a single absorption parameter.

\section{Acknowledgements}

Funding for this work was provided by National Science Foundation grant IIS-1011919 and by a gift from Autodesk. We would also like to thank Kate Salesin and Lydia Wang for their help in preparing and photographing the wood samples, and Fran\c{c}ois Guimbreti\`{e}re for providing lab space for preparing the wood samples.

\appendix

\section{Comprehensive plugin graph}

Parameters are in \textit{italics}, references to volumes in \textbf{boldface}.

\subsection{BSDFs}
Smooth or rough coating BSDF
\begin{itemize}
	\item Bump map BSDF
	\begin{itemize}
		\item Bump height
		\begin{itemize}
			\item \textit{Scale}
			\item \textbf{Pore mask}
		\end{itemize}
		\item BSDF: Sum BSDF
		\begin{itemize}
			\item Diffuse BSDF
			\begin{itemize}
				\item \textbf{Diffuse color}
			\end{itemize}
			\item Alpha blend BSDF
			\begin{itemize}
				\item BSDF 0, Longitudinal wood BSDF: \textbf{Fiber color, highlight width, longitudinal fiber direction}
				\item BSDF 1, Radial wood BSDF: \textbf{Fiber color, highlight width, radial fiber direction}
				\item Alpha: \textbf{Ray mask}
			\end{itemize}
		\end{itemize}
	\end{itemize}
\end{itemize}

\subsection{Distortion}

Distortion may be applied in each of the three cylindrical directions.

\begin{itemize}
	\item \textit{Kernel size}
	\item \textit{Magnitude}
	\item \textit{Impulse size}
	\item \textit{Number of frequency bands}
	\item \textit{Frequency dropoff}
\end{itemize}

\subsection{Volumes}

Diffuse color: Beer's law
\begin{itemize}
	\item \textit{Absorption coefficients} (map from channels/wavelengths to mean absorption length)
	\item \textbf{Path length}
\end{itemize}
Fiber color: Beer's law
\begin{itemize}
	\item Absorption coefficients as diffuse times a \textit{scale} (to make fiber specular less saturated than the diffuse)
	\item \textbf{Path length}
\end{itemize}
Highlight width: Rectangle wave based on time volume
\begin{itemize}
	\item \textit{Minimum value} (how wide is the earlywood highlight?)
	\item \textit{Maximum value} (how wide is the latewood highlight?)
	\item \textit{Proportions of low, rise, high, and fall times per year}
\end{itemize}
Longitudinal fiber direction volume
\begin{itemize}
	\item Extract from \textbf{frame field}
\end{itemize}
Radial fiber direction
\begin{itemize}
	\item Extract from \textbf{frame field}
\end{itemize}
Frame field
\begin{itemize}
	\item Rotate cylindrical frame around $r$ axis
	\begin{itemize}
		\item Angle: Lookup $r$ coordinate into 1D noise
		\begin{itemize}
			\item Wyvill kernel
			\item One band generally sufficient
			\item \textit{Magnitude}
			\item \textit{Kernel width}
		\end{itemize}
	\end{itemize}
\end{itemize}
Path length: sum of volumes
\begin{itemize}
	\item \textbf{Pore mask} times a \textit{scale} (to make pores darker)
	\item Rectangle wave based on \textbf{time volume}
	\begin{itemize}
		\item \textit{Minimum value} (how dark is the earlywood?)
		\item \textit{Maximum value} (how dark is the latewood?)
		\item \textit{Proportions of low, rise, high, and fall times per year}
	\end{itemize}
\end{itemize}
Pore mask: 3D noise volume using bump kernel
\begin{itemize}
	\item Wyvill kernel
	\item Align using \textbf{frame field}
	\item Scale kernel using \textit{aspect ratio} and \textbf{ring porosity}
	\item Single-band
	\item \textit{Pore size}
\end{itemize}
Ray mask: 3D noise volume using bump kernel
\begin{itemize}
	\item Wyvill kernel
	\item Align using \textbf{frame field}
	\item Scale kernel using \textit{aspect ratio}
	\item Single-band
	\item \textit{Pore size}
\end{itemize}
Time volume
\begin{enumerate}
	\item Start with $r$ coordinate
	\item Divide by mean yearly width
	\item Add triangle wave to get $t$
	\begin{itemize}
		\item \textit{Fall slope} (how fast is the growth in the growing season?)
		\item \textit{Rise slope} (how fast is the growth in the off-season?)
		\item \textit{Transition lengths} (how sharply does the growth accelerate?)
		\item Scale so that the median growth rate is the same as the mean
	\end{itemize}
	\item Distort using 1D noise
	\begin{itemize}
		\item Wyvill kernel
		\item One band generally sufficient
		\item \textit{Magnitude}
		\item \textit{Kernel width}
	\end{itemize}
\end{enumerate}
Ring porosity: Rectangle wave based on \textbf{time volume}
\begin{itemize}
	\item \textit{Maximum value} (how large are the pores in the earlywood?)
	\item \textit{Minimum value} (how large are the pores in the latewood?)
	\item \textit{Proportions of low, rise, high, and fall times per year}
\end{itemize}


\begin{figure*}[p]
	\centering
	\includegraphics[width=0.8\linewidth]{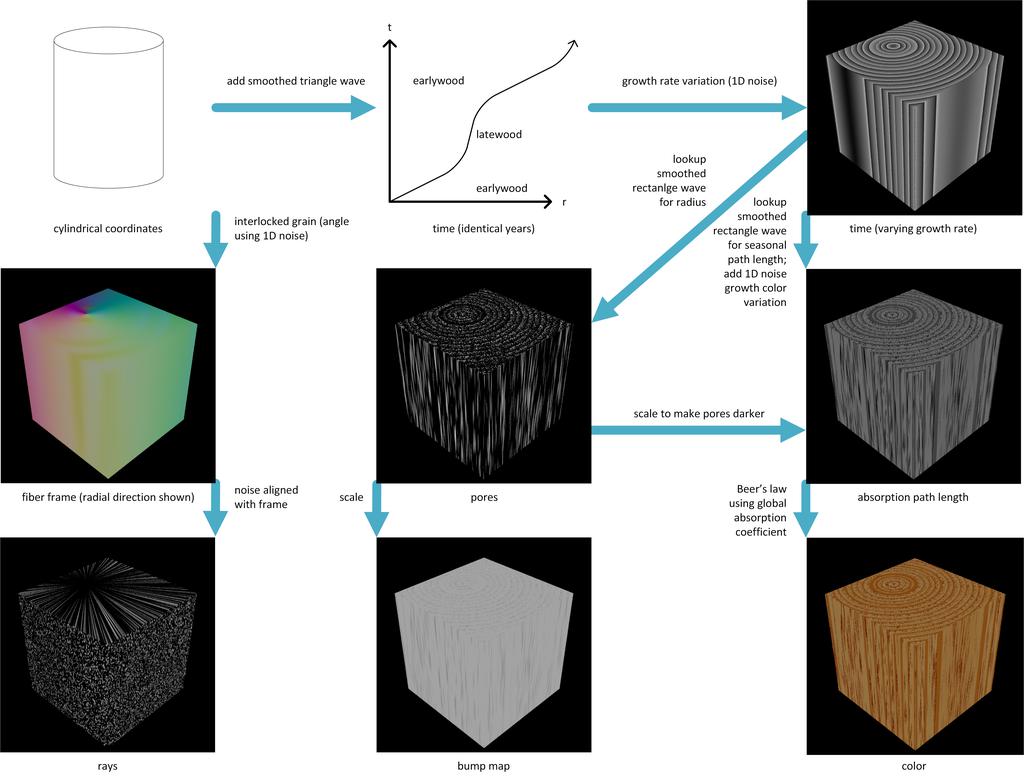}
	\caption{Nodes in the volume texture DAG before distortion is applied. Parameters enter the system at the labeled arrows. All results here are distorted using the same function before the remaining nodes (Figure \ref{fig:post_distortion}) are processed. \albert{Still need to update this.}}
	\label{fig:pre_distortion}
\end{figure*}

\begin{figure*}[p]
	\centering
	\includegraphics[width=0.9\linewidth]{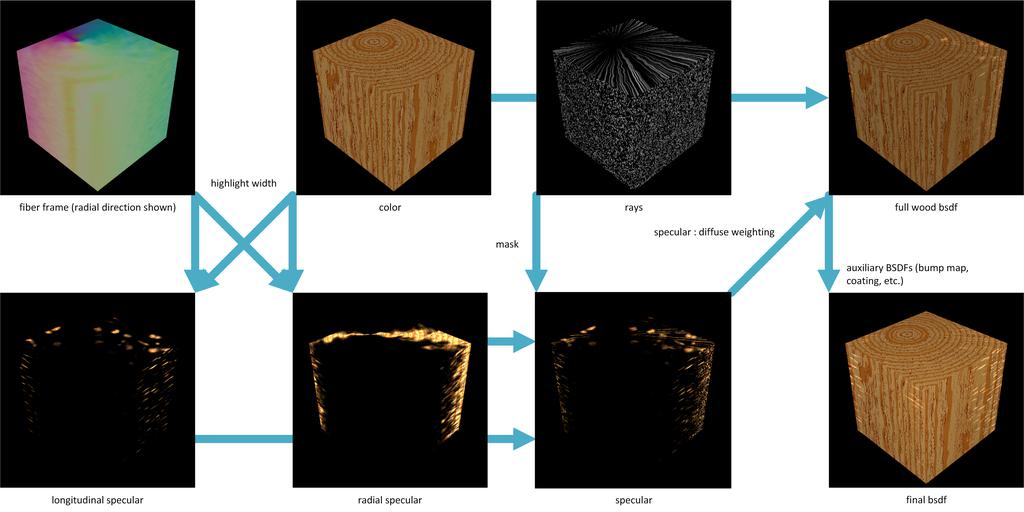}
	\caption{Nodes in the volume texture DAG after distortion is applied. Again, all parameters enter the system at the labeled arrows. As in Figure \ref{fig:pre_distortion} }
	\label{fig:post_distortion}
\end{figure*}

\begin{figure*}[p]
	\centering
	\includegraphics[height=0.35\textheight]{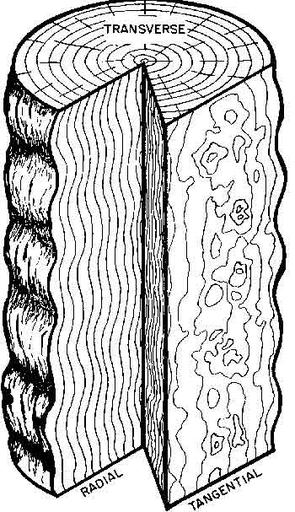} 
	\hspace{0.5in}
	\includegraphics[height=0.35\textheight]{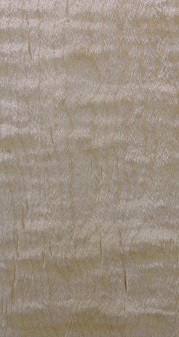}
	\hspace{0.5in}
	\includegraphics[height=0.35\textheight]{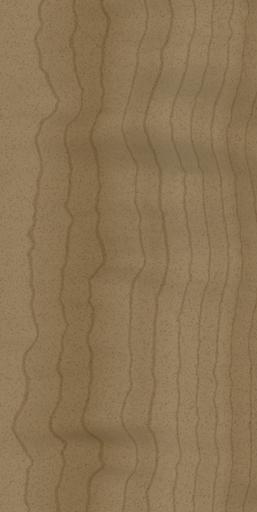} 
	\caption{Distortion in the radial direction produces variation in the shape of growth rings and is responsible for types of figure such as blister and quilted. \textbf{Left:} Diagram of radial waves courtesy of Alabama Agricultural Experiment Station \cite{Beals:1977:FigureInWood}. \textbf{Center:} The ripples in this near-tangential cut of tiger maple are correlated with the shape of the growth rings, indicating that the ripples bend in the radial direction. \textbf{Right:} A rendering showing distortion in the radial direction (only).}
	\label{fig:distortion_radial}
\end{figure*}

\begin{figure*}[p]
	\centering
	\includegraphics[height=0.35\textheight]{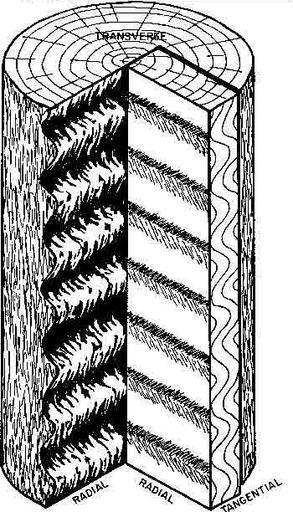}
	\hspace{0.5in}
	\includegraphics[height=0.35\textheight]{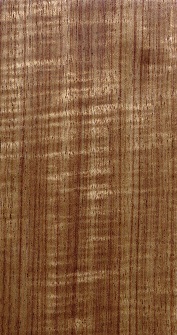}
	\hspace{0.5in}
	\includegraphics[height=0.35\textheight]{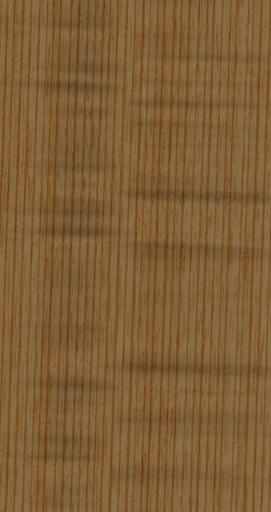}
	\caption{Distortion in the tangential direction does not affect the shape of growth rings, but is responsible for curly figure. \textbf{Left:} Diagram of tangential waves courtesy of Alabama Agricultural Experiment Station \cite{Beals:1977:FigureInWood}. \textbf{Center:} The ripples in this radial cut of canarywood have little effect on the shape of the growth rings, indicating that the ripples bend in the tangential direction. \textbf{Right:} A rendering showing distortion in the tangential direction (only).}
	\label{fig:distortion_tangential}
\end{figure*}

\bibliographystyle{abbrv}
\bibliography{submission} 

\begin{thebibliography}{10}

\bibitem{Barnes:2014:ExactBoundingBoxes}
T.~Barnes.
\newblock Exact bounding boxes for spheres/ellipsoids.
\newblock http://tavianator.com/exact-bounding-boxes-for-spheres-ellipsoids/,
  jun 2014.

\bibitem{Beals:1977:FigureInWood}
H.~O. Beals and T.~C. Davis.
\newblock Figure in wood: an illustrated review.
\newblock Technical Report 486, Alabama Agricultural Experiment Station,
  Auburn, Alabama, USA, Jan. 1977.

\bibitem{Cook:1984:ShadeTrees}
R.~L. Cook.
\newblock Shade trees.
\newblock In {\em Proceedings of the 11th Annual Conference on Computer
  Graphics and Interactive Techniques}, SIGGRAPH '84, pages 223--231, New York,
  NY, USA, 1984. ACM.

\bibitem{Cook:2005:WaveletNoise}
R.~L. Cook and T.~DeRose.
\newblock Wavelet noise.
\newblock In {\em ACM SIGGRAPH 2005 Papers}, SIGGRAPH '05, pages 803--811, New
  York, NY, USA, 2005. ACM.

\bibitem{Galerne:2012:GaborExample}
B.~Galerne, A.~Lagae, S.~Lefebvre, and G.~Drettakis.
\newblock Gabor noise by example.
\newblock {\em ACM Trans. Graph.}, 31(4):73:1--73:9, July 2012.

\bibitem{Hoadley:1980:UnderstadingWood}
R.~B. Hoadley.
\newblock {\em Understanding wood: a craftsman's guide to wood technology}.
\newblock Taunton Press, Newtown, Connecticut, USA, 1980.

\bibitem{Jakob:2014:Mitsuba}
W.~Jakob.
\newblock Mitsuba renderer.
\newblock http://www.mitsuba-renderer.org, 2010.

\bibitem{Kopf:2007:SolidFrom2D}
J.~Kopf, C.-W. Fu, D.~Cohen-Or, O.~Deussen, D.~Lischinski, and T.-T. Wong.
\newblock Solid texture synthesis from 2d exemplars.
\newblock In {\em ACM SIGGRAPH 2007 Papers}, SIGGRAPH '07, New York, NY, USA,
  2007. ACM.

\bibitem{Lagae:2005:ObjectDistribution}
A.~Lagae and P.~Dutr{\'e}.
\newblock A procedural object distribution function.
\newblock {\em ACM Trans. Graph.}, 24(4):1442--1461, Oct. 2005.

\bibitem{Lagae:2008:TileMethods}
A.~Lagae, C.~S. Kaplan, C.-W. Fu, V.~Ostromoukhov, and O.~Deussen.
\newblock Tile-based methods for interactive applications.
\newblock In {\em ACM SIGGRAPH 2008 Classes}, SIGGRAPH '08, pages 93:1--93:267,
  New York, NY, USA, 2008. ACM.

\bibitem{Lagae:2010:SurveyProceduralNoise}
A.~Lagae, S.~Lefebvre, R.~Cook, T.~DeRose, G.~Drettakis, D.~S. Ebert, J.~Lewis,
  K.~Perlin, and M.~Zwicker.
\newblock A survey of procedural noise functions.
\newblock In {\em Computer Graphics Forum}, volume~29, pages 2579--2600. Wiley
  Online Library, 2010.

\bibitem{Lagae:2009:SparseGabor}
A.~Lagae, S.~Lefebvre, G.~Drettakis, and P.~Dutr{\'e}.
\newblock Procedural noise using sparse gabor convolution.
\newblock In {\em ACM SIGGRAPH 2009 Papers}, SIGGRAPH '09, pages 54:1--54:10,
  New York, NY, USA, 2009. ACM.

\bibitem{Lagae:2011:ImprovingGabor}
A.~Lagae, S.~Lefebvre, and P.~Dutr{\'e}.
\newblock Improving gabor noise.
\newblock {\em IEEE Transactions on Visualization and Computer Graphics},
  17(8):1096--1107, 2011.

\bibitem{Lefebvre:2007:RuntimeTextureSynthesis}
S.~Lefebvre.
\newblock Part iv: Runtime texture synthesis.
\newblock In {\em ACM SIGGRAPH 2007 Courses}, SIGGRAPH '07, New York, NY, USA,
  2007. ACM.

\bibitem{Lewis:1984:TextureSynthesis}
J.-P. Lewis.
\newblock Texture synthesis for digital painting.
\newblock In {\em ACM SIGGRAPH Computer Graphics}, volume~18, pages 245--252.
  ACM, 1984.

\bibitem{Lewis:1989:AlgorithmsForSolidNoise}
J.-P. Lewis.
\newblock Algorithms for solid noise synthesis.
\newblock {\em ACM SIGGRAPH Computer Graphics}, 23(3):263--270, 1989.

\bibitem{Marschner:2005:FinishedWood}
S.~R. Marschner, S.~H. Westin, A.~Arbree, and J.~T. Moon.
\newblock Measuring and modeling the appearance of finished wood.
\newblock In {\em ACM SIGGRAPH 2005 Papers}, SIGGRAPH '05, pages 727--734, New
  York, NY, USA, 2005. ACM.

\bibitem{Panshin:1970:TextbookOfWood}
A.~J. Panshin and C.~De~Zeeuw.
\newblock {\em Textbook of wood technology}.
\newblock McGraw-Hillm Inc., New York, New York, USA, 3rd edition, 1970.

\bibitem{Peachey:1985:SolidTexturing}
D.~R. Peachey.
\newblock Solid texturing of complex surfaces.
\newblock In {\em Proceedings of the 12th Annual Conference on Computer
  Graphics and Interactive Techniques}, SIGGRAPH '85, pages 279--286, New York,
  NY, USA, 1985. ACM.

\bibitem{Perlin:1985:PerlinNoise}
K.~Perlin.
\newblock An image synthesizer.
\newblock In {\em Proceedings of the 12th Annual Conference on Computer
  Graphics and Interactive Techniques}, SIGGRAPH '85, pages 287--296, New York,
  NY, USA, 1985. ACM.

\bibitem{Perlin:2002:ImprovingNoise}
K.~Perlin.
\newblock Improving noise.
\newblock In {\em Proceedings of the 29th Annual Conference on Computer
  Graphics and Interactive Techniques}, SIGGRAPH '02, pages 681--682, New York,
  NY, USA, 2002. ACM.

\bibitem{Ramanarayanan:2007:ConstrainedTextureSynthesis}
G.~Ramanarayanan and K.~Bala.
\newblock Constrained texture synthesis via energy minimization.
\newblock {\em Visualization and Computer Graphics, IEEE Transactions on},
  13(1):167--178, 2007.

\bibitem{Shirley:2009:Fundamentals}
P.~Shirley and S.~Marschner.
\newblock {\em Fundamentals of Computer Graphics}.
\newblock A. K. Peters, Ltd., Natick, MA, USA, 3rd edition, 2009.

\bibitem{vanWijk:1991:SpotNoise}
J.~J. van Wijk.
\newblock Spot noise texture synthesis for data visualization.
\newblock In {\em Proceedings of the 18th Annual Conference on Computer
  Graphics and Interactive Techniques}, SIGGRAPH '91, pages 309--318, New York,
  NY, USA, 1991. ACM.

\bibitem{Ward:1992:Anisotropic}
G.~J. Ward.
\newblock Measuring and modeling anisotropic reflection.
\newblock In {\em ACM SIGGRAPH Computer Graphics}, volume~26, pages 265--272.
  ACM, 1992.

\bibitem{Wei:2009:STARExampleBased}
L.-Y. Wei, S.~Lefebvre, V.~Kwatra, G.~Turk, et~al.
\newblock State of the art in example-based texture synthesis.
\newblock In {\em Eurographics 2009, State of the Art Report, EG-STAR}, pages
  93--117, 2009.

\bibitem{Worley:1996:CellNoise}
S.~Worley.
\newblock A cellular texture basis function.
\newblock In {\em Proceedings of the 23rd Annual Conference on Computer
  Graphics and Interactive Techniques}, SIGGRAPH '96, pages 291--294, New York,
  NY, USA, 1996. ACM.

\end{thebibliography}

\end{document}